\documentstyle[epsfig,astrobib,onecolumn]{mn-ab}

\newcommand{\bm}{\bmath}
\newcommand{\bnab}{\mbox{\boldmath $\nabla$}}	
\newcommand{\btimes}{\mbox{\boldmath $\times$}}
\newcommand{\beq}{\begin{equation}}
\newcommand{\eeq}{\end{equation}}
\newcommand{\bmin}[1]{\begin{minipage}[b]{#1\linewidth}}
\newcommand{\emin}{\end{minipage}}

\newcommand{\bfig}   {\begin{figure*}}
\newcommand{\efig}   {\end{figure*}}

\title[Accretion disc--stellar magnetosphere interaction]
{Accretion disc--stellar magnetosphere interaction: field line
inflation and the effect on the spin-down torque.} 
\author[Vasso Agapitou and John C. B. Papaloizou]
       {Vasso Agapitou\thanks{V.Agapitou@qmw.ac.uk} and 
         John C. B. Papaloizou \\
       Astronomy Unit, School of Mathematical Sciences,
       Queen Mary \& Westfield College, London E1 4NS, UK }

\begin{document}

\maketitle

\label{firstpage}

\begin{abstract}
We  calculate  the structure of a force--free magnetosphere which
is assumed to corotate with a central star and which interacts 
with an embedded differentially rotating accretion disc. 
The magnetic and rotation axes are aligned and the stellar field is
assumed to be a dipole.
We concentrate on the case when the amount of field line twisting
through the disc--magnetosphere interaction is {\it large} and consider
different outer boundary conditions. In general the field line
twisting produces field line inflation
(eg. \citeNP{bardou}) and in some cases with large twisting
many field lines can become open.  
We calculate the spin--down torque acting between the star and
the disc and we find that it decreases significantly for cases with
large field line twisting. This suggests that the oscillating torques
observed for some accreting neutron
stars could be due to the magnetosphere varying between
states with low and high field line inflation. Calculations of the
spin evolution of T Tauri stars may also have to be
revised in light of the significant effect that field line twisting
has on the magnetic torque resulting from star--disc interactions.
\end{abstract}


\section{Introduction}
\label{chap5:intro}
 
Accreting magnetised stars  exist in the form of neutron 
stars in close binary systems and T Tauri stars  with
surrounding  protostellar
discs. In both cases an accretion disc is disrupted
through interaction with the stellar magnetosphere
and the accretion flow becomes channelled along 
field lines. Torques act between the star and the accreting
material and they determine the spin history of the star
and its equilibrium rotation period.
A spin--up torque is produced by rapidly rotating
material in the inner regions of the disc
while a spin--down torque is produced by the more
slowly rotating material in the outer parts.
Calculation of the torques requires knowledge of the structure
of the disc and the magnetosphere in which it is embedded.

Early work (\citeNP{ghoshl79a}; \citeNP{ghoshl79b})
suggested a direct dependence of the total
torque on the mass accretion rate.
However, recent observations of accreting neutron stars
(\citeNP{nelson97}; \citeNP{chakra97}) suggest that
the torques can oscillate producing alternating
spin--up and spin--down phases with no evidence of a consistent
correlation between the torque and the mass accretion rate. 
\citeN{liw98} suggest that this may be due
to a variable outer magnetosphere which causes a 
corresponding variation in the spin--down torque
acting between the star and the disc. In this paper we investigate the
structure of a magnetosphere corotating with the central star. We
calculate the spin--down torque due to the star--disc interaction and we find that it is very sensitive
to the amount of field line twisting that occurs through the coupling
of the stellar field lines to the disc. The magnetospheric structure
and the resulting torque are also sensitive to the applied boundary conditions.
In the present work we neglect any torque that may arise from a
stellar wind. 

According to the Ghosh \& Lamb model the magnetic field
of the central star penetrates the accretion disc due to diffusion arising
from the growth of various instabilities. In the outer parts of the
disc, viscous stresses are more important than magnetic stresses so
that the effects of the magnetic field on the disc structure can be ignored.
In the inner parts, the flow becomes dominated by magnetic stresses which
eventually lead to the truncation of the disc at an inner radius,
$R_{\rm d,}$ and the channeling of the flow to the star along
stellar magnetic field lines. The twisting of the poloidal field by
the vertical shear is  assumed to be counteracted by the reconnection of the
oppositely directed toroidal fields through the disc vertical extent.
This occurs on a rapid timescale resulting to an equilibrium ratio for 
toroidal, $B_\varphi$, over vertical, $B_z$ field components at the
disc surface. Alternatively the growth of $B_\varphi$ can be
controlled by turbulent diffusion through the disc
(\citeNP{campbell87}; \citeNP{yi94}) or reconnection of field lines in
the magnetosphere (i.e. \citeNP{liviop}). In all these cases magnetic
torques of a similar form are produced \cite{wang95}. 
 
In all calculations of the magnetic torques resulting from
disc-star interactions $B_z$ is often assumed to be
approximately equal to the stellar dipole field. However, twisting
magnetic field lines imply the existence of currents which will be
distributed in the magnetically dominated magnetosphere which is
assumed to corotate with the central star.  The star-disc
magnetosphere is expected to attain a force--free equilibrium
in an Alfv{\'e}n wave crossing time, as is the case in the solar
corona. In the  force--free equilibrium,
the poloidal magnetic 
field will deviate from that of a dipole, the effect being 
especially significant
in the case of large twisting (or $|B_\varphi/B_z|$). 
We find that the magnetic torque acting on the disc may also be
significantly reduced.

The  related problem of the evolution of an initially potential 
field in response to an increasing footpoint shear through
a sequence of quasi--static force--free equilibria  has been addressed
in the solar context. But note that, unlike in the work presented here,
  ideal
MHD is assumed. Numerical calculations supported theoretical
expectations (\citeNP{aly84}; \citeNP{aly85}; \citeNP{aly91};
\citeNP{sturrock91}) that the energy of the force--free field increases
monotonically
with increasing shear, approaching the energy of an open field
(\citeNP{klimchuck90}; \citeNP{roumeliotis94}; \citeNP{mikicl94}).
Analytical sequences of quasi--static equilibria were also
constructed using a self--similar approach
(\citeNP{nnl92}; \citeNP{lb94}; \citeNP{wolfson95}) and it was
shown that the complete opening of the initially closed field lines is
possible at a finite shear. This is associated with the development of
current sheets as was conjectured by \citeN{aly85}.

\citeN{bardou} have  calculated  magnetospheric  force--free
equilibria  resulting from the 
twisting of an initially dipolar field anchored on a star
which threads an embedded differentially rotating disc.
Disc resistivity  was taken to be of  turbulent origin. 
However, because of problems associated with their
numerical method these authors were not able to calculate equilibrium
configurations for $|B_\varphi/B_z|$  larger than  $2.55$
at the surface of the disc.
In most cases a
maximum $|B_\varphi/B_z| \approx 1$ was adopted.
Nonetheless significant field line inflation was observed.

In the work presented here,  we similarly find that  moderate
to large shearing
of the stellar   dipole field through interaction with the
disc can lead to a partially open configuration, with expulsion
of toroidal field,  for
favourable outer boundary conditions. Although the process
can be inhibited by the presence of an outer boundary that
confines the field, significant field line inflation may
occur with a large reduction of the magnitude of the vertical
field in comparison to the initial dipole value. This could have
important consequences for studies of stellar spin evolution which
have assumed a dipolar poloidal field at equilibrium
(\citeNP{ghoshl79b}; \citeNP{cameron93}; \citeNP{yi95}; \citeNP{ghosh95};
\citeNP{armitage96}). 
 
After formulating the physical model and basic equations
in Section~\ref{eqs}, we describe our numerical method in
Section~\ref{chap5:numeth}. Using this we were able to perform
calculations with a maximum value of $|B_\varphi/B_z|$ at the disc surface
 of $120$ and with a variety of external disc radii.
Results for Keplerian discs with different
amounts of field twisting are given in sections~\ref{sec:chap4numres}.
These calculations were performed with a field--confining, perfectly conducting outer boundary.

We also
performed calculations with a partly sub--Keplerian
accretion disc 
using the form of angular velocity, $\Omega$ given by \citeN{campbell87}. This emulates
the existence of an inner boundary layer as  proposed in
the Ghosh \& Lamb model. Results are given in
Section~\ref{subsec:newom} for the conducting outer boundary
condition. In Section~\ref{subsec:secbc}
 we investigate the effect of the outer boundary condition,
 showing that a boundary condition that allows field lines to
penetrate it, leads to more open configurations.

In Section~\ref{subsec:torque}
 we go on to evaluate the spin down torque
acting  between star and disc for the force free configurations.
We show that this torque can be reduced by up to
a factor of $100,$ in comparison to what
would be obtained assuming an unmodified  dipole field,
  in a configuration that has undergone
large twisting and toroidal field expulsion.

Finally in Section~\ref{Discussion} we summarize and
discuss our results and their application.

\section{Basic Equations}
\label{eqs}
\subsection{Force--free equilibria}

\noindent We consider a  magnetic field ${\bm B_{\rm dp}}$ anchored on
a star that rotates with a rate $\Omega_{\star}.$ We suppose that the
magnetic  field is axisymmetric with  symmetry axis aligned with the
rotation axis. A  thin accretion disc orbits in the equatorial plane 
such that the system  remains axisymmetric.
We  ignore the internal dynamics of the disc and assume  that  the
material rotates with a prescribed angular velocity $\Omega.$
 For the most part this will equal or be close to
the local Keplerian  angular velocity
$\Omega_{\rm K} \equiv \sqrt{GM/R^3}$ where $R$ is the radial distance
to the star  measured in the midplane of the disc. We suppose that there
is a highly conducting low density corona that corotates with the star
and in which the disc is immersed.

However, the disc has non zero resistivity such that the stellar field
is enabled to diffuse into the disc. Because the star/corona  and disc
rotate at different rates, field lines that permeate the disc are
twisted  in the azimuthal direction such that a toroidal component of
the magnetic field is generated. 

The disc is  considered to be  turbulent and 
 manifesting an effective  magnetic diffusivity,
$\eta.$  
This enables a steady state to be achieved in which the
generation of toroidal  magnetic field  through differential rotation
between the disc and star/corona is balanced by  the effects of 
turbulent diffusion. 

The corona responds to the twisting of the initially dipolar field by 
producing toroidal magnetic field  and current density
components. These adjust such that a force--free equilibrium is
eventually attained. We expect  this situation to occur when the
coronal Alfv{\'e}n speed sufficiently exceeds the characteristic
rotational speeds. This  requires a low density corona which is
magnetically dominated. For the purposes of the work presented here,  we 
assume that such a corona exists in which a force--free equilibrium  is
established in a time short compared to the stellar rotation
period. Thus we look for steady state axisymmetric  configurations
involving the star, the disc and the corona. 

An axisymmetric  
magnetic field ${\bm B}$ can be split into poloidal ${\bm B}_{\rm p}$
and toroidal $B_\varphi {\hat \bvarphi}$   components such that:
 
\beq  {\bm B}_{\rm p} = \bnab \btimes (\Psi {\hat \bvarphi} /(r\sin\theta)),
 \eeq  

\noindent where $\Psi$ 
is the magnetic stream function which is constant
on field lines. Here we use spherical polar coordinates $(r,\theta,\varphi)$
 based on the central star. Then
 
\beq
 {\bm B}_{\rm p} =\left  ( \frac{1}{r^2 \sin{\theta}}\frac{\upartial \Psi}{\upartial
 \theta},
 -\frac{1}{r \sin{\theta}}\frac{\upartial \Psi}{\upartial r}\right)  \label{BRS}\eeq

The force--free condition implies that the  Lorentz force is zero
everywhere, thus: 

\beq
{\bm J} \btimes {\bm B} = 0
\label{FORCEFC}
\eeq

\noindent where ${\bm J} = (c/4\upi) \nabla\times {\bm B}$ is the
current density (c.g.s units are used throughout). The toroidal
component of equation (\ref{FORCEFC}) gives:

\beq
{\bm B}_{\rm p} \cdot \bnab (r \sin{\theta} B_\varphi) = 0.
\label{BPHI0}
\eeq

\noindent The definition of $\Psi$ implies constancy on field lines
  or equivalently ${\bm B}_{\rm p} \cdot \bnab \Psi = 0$. Thus
(\ref{BPHI0}) implies that:

\beq
r \sin{\theta} B_\varphi = f(\Psi) \equiv f,
\label{FCOND}
\eeq

\noindent where $f$ is an arbitrary function of $\Psi.$

Using (\ref{FCOND}) and the poloidal component of (\ref{FORCEFC}) we
finally obtain the governing equation for $\Psi$ in the form:

\beq
\Delta \Psi = -f f'
\label{EQN}
\eeq

\noindent where $f' \equiv {\rm d}f/{\rm d} \Psi$ and $\Delta$ is the
differential operator defined through:

\beq
\Delta = \frac{\upartial^2 }{\upartial r^2} + \frac{1-\mu^2}{r^2}
\frac{\upartial^2 }{\upartial \mu^2} ,
\label{EQN0}
\eeq

\noindent where $\mu \equiv \cos{\theta}$.

\noindent Equation (\ref{EQN}) can be solved for $\Psi$ if $f$ 
is known. Since $f(\Psi)$ is constant on poloidal field lines
 we need to calculate its value at only one point along each 
field line. This we do by consideration of the interaction
of the field with the resistive disc.

\subsection{Disc--stellar field interaction}
\label{subsec:discfi}

The corona above the disc is assumed to
corotate with the star and so a toroidal field will be generated due
to the vertical shear. 
 Neglecting the radial component of the field interior to the disc,
we calculate the internal toroidal field  using 
the toroidal
component of the induction equation which in cylindrical coordinates
$(R,\varphi,z)$ takes the form:

\beq
\frac{\upartial B_{\varphi}}{\upartial t} = R {\bm B}_{\rm p} \cdot
\nabla \Omega + \frac{\eta}{R} \Delta_{\rm c} (R B_\varphi) + \frac{1}{R}
\nabla \eta \cdot \nabla (R B_\varphi)
\label{INDTOR}
\eeq

\noindent where $\Delta_{\rm c} = \nabla^2 -(2/R)(\upartial/\upartial
R)$ and $\eta = {\mathcal D} \nu$ is
the turbulent magnetic diffusivity  which is
assumed to be proportional to the turbulent viscosity $\nu,$
the constant of proportionality being ${\mathcal D}.$ 
We use  the Shakura \& Sunyaev (1973)  parameterization
for $\nu,$
$\nu =\alpha_{\rm SS} \Omega_{\rm K} H^2.$
Here the viscosity parameter is  $\alpha_{\rm SS}$ and 
$H$ is the disc semithickness. The  gas velocity in the disc is taken to be 
${\bm u} = (0,r \Omega,0).$

We consider that $B_\varphi$ is generated by the shearing of the
vertical magnetic field component  arising from the $z$ dependence
of $\Omega.$  For simplicity we take $\eta$ to be 
a function of radial distance only. For a more 
general treatment see \citeN{campbell87}. Noting that the
disc is thin we expect $\upartial/\upartial z \sim (R/H)
\upartial/\upartial R$ and so we neglect radial 
derivatives in equation (\ref{INDTOR}). Then in a steady state we have:

\beq
\eta  \frac{\upartial^2 B_\varphi}{\upartial z^2}  = - R B_z \frac{\upartial
\Omega}{\upartial z}
\label{INDTF}
\eeq
To find $B_\varphi$ 
we need to specify $\upartial \Omega/\upartial z$. 
We follow previous authors (\citeNP{wang87}; \citeNP{campbell87};
\citeNP{yi94}) and take the vertical shear to be concentrated in a
thin layer close to the disc surface. There the  angular velocity changes
 between the  disc midplane value  and $\Omega_\star.$ We take the vertical average
of equation (\ref{INDTF}) the right hand of which  gives:

\beq
-{RB_z \over H} \int_0^H\frac{\upartial\Omega}{\upartial z} dz = -\frac{RB_z(\Omega_\star -
\Omega)}{ H},
\label{VSHEAR}
\eeq

\noindent where $B_z$ is  assumed not to  vary with
$z$ thus  being the disc midplane value. In the following $\Omega$
will refer to the disc midplane value. 

\noindent We assume a simple approximation for the vertical average of the  dissipative term on  the
left-hand-side of equation (\ref{INDTF}) namely:

\beq
{1\over H} \int_0^{H} \eta \frac{\upartial^2 B_\varphi}{\upartial z^2}dz \approx -\eta\gamma
\frac{B_\varphi}{H^2},
\label{BPHIDF}
\eeq

\noindent where $B_\varphi$ now applies to the value at the upper disc
surface and $\gamma$ is a dimensionless parameter  expected to be of
order unity. 

Combining equation (\ref{VSHEAR}) with
equation (\ref{BPHIDF}) we obtain the following
estimate for the equilibrium value of the ratio $B_\varphi/B_z$ 
at the {\it upper} disc surface:

\beq
\frac{B_\varphi}{B_z} =  \pm\frac{R H}{\gamma \eta}(\Omega_\star -
\Omega)
\label{BPHIBZG}
\eeq
We have inserted the $\pm$ alternative to allow
for different senses of rotation of the star and disc
with respect to the  right handed coordinate system with the
negative sign corresponding to clockwise rotation.
The angular velocities are then always positive.
Using  the turbulent diffusivity specified above
and taking  $\gamma=1,$  
equation (\ref{BPHIBZG}) gives for $\Omega=\Omega_K$:

\beq
\frac{B_\varphi}{B_z} = \pm \frac{R}{{\mathcal D}\alpha_{\rm SS}
H}\left[ \left(\frac{R}{R_{\rm c}}\right)^{3/2} - 1 \right]
\label{BPHIBZ0}
\eeq

From equation (\ref{BPHIBZ0}) we see that $B_\varphi$ changes 
sign at  the corotation radius, $R_{\rm c} \equiv (G
M/\Omega^2_\star)^{1/3}$. Similarly
the vertically integrated $z \varphi$
component of the magnetic torque per unit area, $B_z B_\varphi R^2/(2\upi)$, also changes
sign such that the star transfers angular momentum to the disc outside the
corotation radius while it gains angular momentum from disc material
inside that radius.

In our calculations we shall assume that $R H \Omega_\star/(\gamma \eta)
\equiv C = \mbox{constant}$. Equation (\ref{BPHIBZ0}) then becomes
(adopting the negative sign corresponding to clockwise rotation):

\beq
\frac{B_\varphi}{B_z} = C \left[\left(\frac{R_{\rm c}}{R}\right)^{3/2}
- 1 \right]
\label{BPHIBZ}.
\eeq

\noindent Our assumption of $C = \mbox{constant}$ is such that for
large values of $R/R_{\rm c},$ $B_\varphi/B_z \approx -C$ which is constant.
However, an important result of the calculations,
namely  the inflation of poloidal field lines
such that  $B_z\rightarrow 0$ at the disc surface
for large $C$, should not depend on having a  dependence
of $C$ on radius  provided it  remains large. For $H/R = 0.1$,
$\alpha_{\rm SS} = 0.01$ and ${\mathcal D} = \gamma = 1$ at the
corotation radius, these being  reasonable values for accretion discs, equation
(\ref{BPHIBZ}) gives $|B_\varphi/B_z| = C  = 10^3$
for $R \gg R_{\rm c}.$ We comment that equation (\ref{BPHIBZ0}) indicates that
in a disc with constant $H/R$ and constant $\alpha_{SS},$

\beq 
C \propto \frac{1}{{\mathcal D}}
 \left(\frac{R}{R_{\rm c}}\right)^{3/2}. 
\eeq

\noindent Thus in this case  if  ${\mathcal D} = 1$, $C$ would
indeed increase with radius.
              
Our analysis is distinct from that of e.g. \citeN{liviop} who use a
similar expression for $|B_\varphi/B_z|$ to equation 
(\ref{BPHIBZ}) (for $R \ge R_{\rm c}$) but who assume that fast
reconnection in the star--disc corona with the coronal Alfv{\'e}n speed
restricts $|B_\varphi/B_z|$ to a maximum value of unity. The above
discussion suggests that $C$ and hence $|B_\varphi/B_z|$ at the disc
surface is large. The actual magnitude of  $|B_\varphi|$ is then
controlled by the requirement of a force free equilibrium in the
corona. This results in a significant departure of the poloidal field
from the original  stellar dipole through a significant inflation and
opening out of the field lines. 

Equation (\ref{EQN}) is solved for $R_{\rm d} < r <
R_{\rm ext}$ where $R_{\rm d}$ and $R_{\rm ext}$ correspond to the inner and
outer disc radii. $R_{\rm d}$ is usually  taken to be small
enough that 
 the magnetic torque that is applied  there  overwelhms the
viscous torque so that angular momentum transport is magnetically
regulated.  

\subsection{The location of the inner disc radius}
\label{subsec:rdin}

An accurate calculation of the inner disc radius would require
the solution of the accretion disc structure problem with the
inclusion of all magnetic
stresses and allowing for sub--Keplerian rotation
in the innermost disc region. In most studies an approximate
value for $R_{\rm d}$ has been calculated based on the
criterion of disc disruption at a radius where magnetic stresses
start to dominate viscous stresses. 
In order for stellar accretion  to
proceed the magnetic field must not 
disrupt the disc  exterior to the corotation radius since the 
magnetic torque for $R > R_{\rm c}$ 
imparts  angular momentum to the 
disc gas  which cannot connect to the stellar field. 
Assuming  accretion takes place so that $R_{\rm d} < R_{\rm c},$
the inner disc radius can be estimated
by requiring that the magnetic torque balances the rate of angular
momentum  advection by the flow or that: 

\beq
\dot{M}\left[ {d \left(\Omega R^2\right)\over dR}
\right]_{R=R_{\rm d}} = \mp [B_\varphi B_z]_{r=R_{\rm d}} R^2_{\rm d}
\label{RDEQ}
\eeq

\noindent where ${\dot M}$ is the mass accretion rate and the positive sign corresponds to clockwise rotation. 
Because of the rapid increase of the magnetic
field strength inwards estimates of
$R_{\rm d}$  using equation (\ref{RDEQ}) (e.g. \citeNP{wang95};
\citeNP{yi95}) give $R_{\rm d}$ close to $R_c$ when
the star is near its equilibrium spin rate.  
 
However, we point out that this 
analysis is based on an assumed  {\it stellar dipole} structure for $B_z$. 
When
the effect of an increasing $B_\varphi$ at the disc surface is
taken into account we expect that both $B_z$ and the  magnetic torques
will be modified. We will
return to this point below.

In our numerical calculations we have mostly set $R_{\rm d} = R_{\rm c}$
implicitly assuming that
the star is close to its equilibrium
spin rate. For completeness 
we  also performed some calculations with a sub--Keplerian inner
disc region (see Section~\ref{subsec:newom}).

\section{Numerical solution}
\label{chap5:numeth}

\subsection{Previous work}

Equation (\ref{EQN}) has been solved by \citeN{bardou} 
using a relaxation method with a coordinate transformation $r \rightarrow 1/r$ 
so that the solution can extend to arbitrary large radii in principle. 
\citeN{bardou} adopted
equation (\ref{BPHIBZ0}) for $B_\varphi/B_z$ with a maximum expected value of $C \sim 3.25 \cdot 10^3 (R_{\rm d}/R)^{1/8}.$ However, in practice the numerical
method would not converge unless $C$ was multiplied by a small
parameter $\lambda$ whose value depends on $R_{\rm c}/R_{\rm d}.$
 As a result the maximum value of $|B_\varphi/B_z|$ at the disc
surface that could be obtained in a converged calculation was $2.55.$
For this particular case $\lambda = 1.6 \cdot 10^{-5}$ (A.Bardou, private communication).
\citeN{bardou} 
 found that the force--free
equilibria consist of closed field lines whose footpoints are shifted
radially outwards with respect to their position in the dipolar state
and with field lines that are flattened at small $\theta$ as compared
to dipolar ones. \citeN{bardou99} 
found  the same indication from a similarity solution.
Numerical convergence
problems were encountered also by \citeN{wolfsonl92} who used a
 relaxation method to solve equation (\ref{EQN}) with a prescribed $f$
in the context of the solar corona. They found that no equilibrium
solution could be calculated for $|B_\varphi / B_z| \ga 1$.
We comment that methods for solving  equation (\ref{EQN})
 based on a simple iteration method for producing
a contraction mapping (see also \citeNP{wolfsonv91})
using successive  previous approximations for the right hand side
tend to fail to converge if $ff'$ is a sufficiently rapidly varying
function of $\Psi.$ By comparison with solutions of related algebraic
problems, the lack of convergence in solving equation (\ref{EQN}) is
not necessarily related to the emergence
of additional solutions through a bifurcation. On the
 other hand studies that employ different methods of solution do not
seem to encounter convergence problems (i.e. \citeNP{klimchuck90};
\citeNP{roumeliotis94}).

We conclude that previous studies of force--free equilibria related to
star--disc interactions were restricted to  modest values of 
$|B_\varphi/B_z|$ at the disc surface due to restrictions enforced by 
the adopted numerical method.

\subsection{Numerical method} 

In this paper we adopt a different method of solution of equation
(\ref{EQN}) to those indicated  above. In our case  equation (\ref{EQN}) can be solved  without the
use of a controlling parameter $\lambda$ and in principle for any
disc surface value of $|B_\varphi/B_z|.$ In practice some restrictions
also apply in this case  but we were able to perform
calculations for  discs with  a ratio of inner to outer radius
of several decades  and for a maximum value
of $|B_\varphi/B_z| \sim 10^2$. 

In our method equation (\ref{EQN}) is transformed into a parabolic 
PDE by moving the source term to the left of equation (\ref{EQN}) and 
by equating the new expression to a  multiple of the time derivative of $\Psi$ which
is expected to tend to zero  close to
an equilibrium configuration. We use an explicit
finite--difference scheme to integrate the parabolic PDE forward in time as an initial value 
problem until a steady state is achieved.
Following this approach rather than solving the elliptic
equation (\ref{EQN})  directly, 
we were able to obtain equilibria corresponding to a wide
range of values of the source term.

The total magnetic stream function, hereafter denoted by $\Psi_{\rm t}$, can be written as a
sum of contributions arising  from  disc currents,  $\Psi_{\rm d}, $
and  the  central dipole, $\Psi_{\rm dp}.$ Thus $\Psi_{\rm t} =
\Psi_{\rm dp} + \Psi_{\rm d}$, where $\Psi_{\rm dp}$ is given by:  

\beq
\Psi_{\rm dp} = - K\frac{1-\mu^2}{r},
\label{PSIDP}
\eeq

\noindent and $K$ is a normalizing constant. Note that 
$\Delta \Psi_{\rm dp}  = 0$ since $ f = 0$ for a dipole field.
Equation (\ref{EQN}) with $f \ne 0$ is then solved for $\Psi_{\rm d}$ only
from:

\beq 
\Delta \Psi_{\rm d} = -f(\Psi_{\rm t})f'(\Psi_{\rm t}).
\label{EQN1}
\eeq 

We  find it convenient to
drop the subscript ``${\rm d}$'' from the disc magnetic stream function  and 
 use dimensionless quantities, ${\bar r} = r/r_\star$, ${\bar \Psi} = \Psi /
\Psi_0$ and ${\bar B} = B/B_0$ where $B$ represents the
magnitude of either the toroidal or the poloidal magnetic field
components and we take $\Psi_0 = B_0 r^2_\star$ such that $B_0 =
B_{\rm dp}(r_\star,0)$, the magnitude
of the dipole field on the equatorial plane. Equations (\ref{EQN}) and
(\ref{EQN1}) can then be read
in dimensionless form if all quantities are replaced by their barred 
equivalents. For simplicity we will drop the bars from the
dimensionless quantities from now on. 

The computational domain is bounded by an inner radius $r_{\rm in}$, 
an outer radius $r_{\rm out} = R_{\rm ext}$, the disc midplane  at $\mu
= 0$ and the symmetry axis, $\mu =1.$ The grid
spacing is taken to be uniform in both $r$ and $\mu.$  

At $r = r_{\rm in}$ and $\mu = 1$ respectively
we specify the boundary condition $\Psi = 0$. The first condition
specifies only the dipole flux exists at the inner boundary while the
second is a requirement of the coordinate system. 
On the disc midplane, $\mu = 0$, we impose the symmetry condition,
$\upartial \Psi/\upartial \mu = 0$. 

We have considered two different outer boundary conditions. The first one
is the Dirichlet condition: $\Psi = 0$ which forces the field due to
the  disc to be tangential there. Physically this corresponds to a
conducting boundary which excludes the field arising from currents in the
disc but not the original dipole field which may be assumed to have
had infinite time to diffuse. The second outer boundary condition that
we use is the Neumman condition: ${\upartial \Psi \over \upartial r}  =
0.$ This makes the disc field radial
at the outer boundary as might be the case if there were a coronal wind there.
The second boundary condition leads to much more open configurations
than the first and hence to larger departures from the stellar dipole field.

We add a term $F(\upartial \Psi/\upartial \tau)$ to the right-hand side
of the dimensionless form of equation (\ref{EQN1}) where
we are allowed to choose $F$ as an arbitrary function of position  and  solve  
the  equation  with a forward in time, $\tau$,  and centered in space (FTCS)
finite--difference scheme on a $(Nr\times Nm)$, $(r,\mu)$ computational grid.   
The finite--difference equation that 
we use is  given  at a grid point denoted with subscript  $(i,j)$ by:

\begin{eqnarray}
F_{i,j}\frac{\Psi_{i,j}^{n+1} - \Psi_{i,j}}{{\rm d}\tau} 
\!\!\!\!&=&\!\!\!\! \frac{\Psi_{i+1,j} - 2 \Psi_{i,j} +
\Psi_{i-1,j}}{({{\rm d}x})^2} \nonumber\\
\!\!\!\!&+&\!\!\!\! \frac{1-\mu^2_j}{r^2_i} \frac{\Psi_{i,j+1} - 2 \Psi_{i,j} +
\Psi_{i,j-1}}{({{\rm d}\mu})^2} +
 f\left(\left.\Psi_{\rm t}\right|_{i,j}\right)f'\left(\left.\Psi_{\rm
t}\right|_{i,j}\right). 
\label{EQF}
\end{eqnarray}

\noindent where ${\rm d}r,$ and  ${\rm d}\mu,$ are the uniform grid
spacings in $r, $ and   $\mu,$  respectively. Variables without superscripts 
are at $\tau$ level $n.$ The time step is ${{\rm d}\tau}.$ We assume
that the disc is truncated at $R_{\rm d}$ by an infinite conductor so we have $B_\varphi = 0$ for
$1< R \le R_{\rm d}.$  For all other values of $R$ on the  equator ($\mu=0$
 denoted by subscript $1$) we use equation (\ref{BPHIBZG})
approximated as follows: 

\beq
\left. B_\varphi \right |_{i+1/2,1} = C \,  \left. B_z \right|_{i+1/2,1}
 \frac{\Omega_{i+1/2} - \Omega_\star}{\Omega_\star}.
\label{BPHIF} 
\eeq

\noindent In equation (\ref{BPHIF}) 
$\Omega_{i+1/2}$ is either given by the Keplerian rate or by the form 
proposed by \citeN{campbell87} (see Section~\ref{subsec:newom}).  
$\Omega_\star$ is calculated on the $(i+1/2,1)$  grid point that is
closest to the prescribed value of $R_{\rm c}$. The vertical magnetic
field at the disc surface is calculated at the $(i+1/2,1)$ points by
numerical differentiation of equation (\ref{BRS}). 

Using equation (\ref{BPHIF}) we calculate
$f_{i+1/2,1} = R_{i+1/2} \left. B_\varphi\right |_{i+1/2,1}$
while $(ff')_{i,1}$ is calculated from:

\beq
(ff')_{i,1} = \frac{1}{2} \frac{f^2_{i+1/2,1} -
 f^2_{i-1/2,1}}{\left.\Psi_{\rm t}\right|_{i+1/2,1} -
 \left.\Psi_{\rm t}\right|_{i-1/2,1}}.
\label{FFDASH}
\eeq

\noindent Since 
$f$ is a function of $\Psi_{\rm t}$ only we construct a table of values 
of $f$ as function of $\left. \Psi_{\rm t}\right|_{i,1}$ on the
equator. We then use this table to calculate $f$ for any  required
value of $\Psi_{\rm t}$ in equation (\ref{EQF}) using linear
interpolation. 

The main disadvantage of the method we use is the constraint on the
time step coming from the   requirements of numerical  stability. 
 We find that 
the maximum allowed timestep
decreases rather steeply with $C.$

In order to achieve a more rapid evolution for the same 
computational time and always keeping in mind that we 
are only interested in final steady state equilibria,
we have introduced a spatially varying ``diffusion coefficient'',
$(1/F)$, that
multiplies the right-hand side of equation (\ref{EQF}).
This coefficient is taken to be  equal to ${\rm d}\tau_{i,j}/ \min({\rm
d}\tau_{i,j})$ and therefore allows for a more advanced 
evolution of the values at the grid points where ${\rm d}\tau_{i,j},$
the maximum allowed timestep based on local stability considerations,
can be larger. Since the source term peaks near the corotation 
radius but it is significantly smaller elsewhere it is possible
to reach the equilibrium in significantly smaller computational
time  in this way.

Time integrations have been performed until the right hand side of
equation (\ref{EQF}) becomes smaller than $\sim 10^{-6}$ 
in dimensionless units although the
magnetic field configuration is usually found to have converged to its
final form well before  this criterion is satisfied.

\section{Numerical Results}
\label{sec:chap4numres}

The parameters used in the calculations presented here are
summarised in Table~\ref{table1}. All the calculations were performed
with $r_{\rm in} = 1$. Different values were
used for $R_{\rm ext}$ and $C$ and calculations were done using 
both outer boundary conditions on $\Psi.$  All the calculations
with $\Omega=\Omega_{\rm K}$ were performed with $R_{\rm
d} = R_{\rm c}$. Our choice of $R_{\rm c} = 3$ implies $\Omega_\star = 0.19 \Omega_{\rm K}(R=R_\star)$
which corresponds to a rotation period of $\approx 5 \,{\rm days}$ for
$R_\star = 3 R_\odot$ and $M_\star = 0.5 M_\odot$. The observed
rotation periods of CTTS are $3$--$12$ days (\citeNP{bouvetal97} and
references therein). 

The choice of $R_{\rm ext}$ was dictated mainly by the limitations of
the computational method since a much larger value of $R_{\rm ext}$
would require increased values of $Nr$ and $Nm$ for the same resolution to be
achieved as in the cases with smaller outer disc radii.

We chose $C=3$--$120$ in order to examine the effect of an extended
range of values of $|B_\varphi/B_z|$  at the
disc surface on the magnetospheric equilibrium. 
Various authors have assumed $|B_\varphi/B_z| \sim 1$ at the disc
surface in their
analysis. \citeN{ghoshl79a} argue that anomalous resistivity in the
disc would  limit $B_\varphi/B_z.$  \citeN{alyk90}  and \citeN{liviop}
assert that fast reconnection outside the disc would lead to a
similar result. These physical arguments have not so far
been supplemented by detailed calculations in the context
of star--disc interactions. Our maximum value of $C$ was chosen
so that the computational times required are reasonable. 
Nevertheless much larger values of $C$ and therefore of $B_\varphi$ on
the disc surface could result in very large toroidal magnetic pressure
on the disc surface and the subsequent destabilisation of the disc.

In Table~\ref{table1} we have also included the parameters of
calculations performed with a sub--Keplerian form of $\Omega$. This
is characterised by the ``fastness'' parameter $\omega \equiv
\Omega_\star/\Omega_{\rm K}(R_{\rm d})$ (for more details see
Section~\ref{subsec:newom}). 

\begin{table}
\begin{center}
\begin{tabular} {cccccccccccccccccccccccccccccccccccccccccc} \hline \hline

B.C.& $C$ & $R_{\rm ext}$ & $\omega$ & $Nr,Nm$ & $|B_\varphi/B_z|_{\rm \mu=0}$ \\
\hline\hline
$1^{\rm st}$ & 3  &11 & 1     & 200,80 &  2.56  \\
             &    &21 &       &        &  2.83  \\
             &    &60 &       &        &  2.96  \\
             &    &80 &       &        &  2.97  \\
\hline
             & 10 &60 &       & 100,50 &  9.85  \\
             &    &80 &       &        &  9.90  \\
\hline
             & 30 &11 &       & 50,30  &  25.11  \\
             & 30 &21 &       & 120,50 &  28.18  \\
             & 30 &60 & 0.875 & 100,50 &  29.60  \\
             & 30 &11 & 0.2   &  50,30 &  23.86  \\
             & 30 &   &       & 100,50 &  24.90  \\
\hline
             & 60 &   & 1     & 50,30  &  50.21  \\
             & 60 &60 & 0.875 & 100,50 &  59.21  \\
\hline
             &120 &11 & 1     & 50,30  &  100.42    \\
             &120 &60 & 0.875 &120,50  &  {\bf 118.30} \\
\hline
\hline
$2^{\rm nd}$ & 3  &60 & 1     & 200,80 &  2.96   \\
             & 10 &   & 0.875 & 120,50 & 9.86   \\
             & 60 &   &       &        & 59.15   \\
             & 120&   &       &        & {\bf 118.30} \\
\hline
\hline
\end{tabular}
\end{center}
\caption{Parameters used in the numerical calculations described
 in the text.
 All calculations
above the third pair of horizontal lines were performed
with the first (Dirichlet) outer boundary condition: $\Psi =0$ at
$r=R_{\rm ext}$. Those below were performed with the second (Neumman)
outer boundary condition which has the normal
derivative of $\Psi$ on $r=R_{\rm ext}$ vanishing. For both outer boundary
conditions the maximum value of $|B_\varphi/B_z|$ in the calculations,
given in the last column, is $118.3$. Additional calculations to those
listed here were performed in order to study the variation of the
torque between the star and the disc with $C.$} 
\label{table1}
\end{table}

\subsection{Qualitative considerations of star--disc interactions}
\label{subsec:Crange}

The main result of the production of toroidal field in the star--disc
corona is the presence of a feedback mechanism 
that reduces $|B_z|$ at the disc surface. In this way
the magnitude of $B_{\varphi}$ does not become
too large to maintain a force free equilibrium when $C$ is very large.
As $C$ is increased, the field becomes increasingly
twisted. The result is the production of azimuthal currents
as well as a toroidal field component in the corona. The azimuthal currents
tend to reduce $|B_z|$ near the disc and hence the magnitude of $B_{\varphi}.$
To see this in a qualitative way: the disc flux obeys equation
(\ref{EQN1}) which, for the boundary conditions used, can be converted to an
integral form:

\beq 
\Psi =\int_V G(r,\mu,r',\mu')f(\Psi_{\rm t})f'(\Psi_{\rm t})
r'^2d\mu' dr'.
\eeq

\noindent Here $G$ is an appropriate Green's function and the stream
function inside the integral, taken over the computational domain,
 is evaluated using the primed coordinates. Thus:

\beq 
\Psi_{\rm t} =\int_V G(r,\mu,r',\mu')f(\Psi_{\rm t})f'(\Psi_{\rm t})
r'^2d\mu' dr' + \Psi_{\rm dp}.
\eeq

\noindent For small $C,$ $f$ is small and $\Psi_{\rm t}$ is close to
the dipole value $\Psi_{\rm dp}.$ However, for large $C$  and hence
$f,$ a solution can only be found (or the force free equilibrium
can only be maintained) by otherwise reducing the magnitude of the integral.
Note that if $\Psi_{\rm dp}$ is scaled by multiplying by a constant
value, both $\Psi_{\rm t}$ and the integral above scale similarly.
The magnitude of the integral  can be reduced by having fewer field
lines intersecting the disc
and therefore more field lines for which $f=0$. If the outer boundary
condition does not allow that, the magnitude of the integral can also
be reduced by the field lines intersecting the disc at progressively
larger radii (and smaller field magnitudes), as $C$ increases.
In either case, the magnitude of $B_{\varphi}$ is controlled
and the poloidal field takes on either a more inflated or a partially
open structure. 

We now discuss our results 
for a disc with Keplerian rotation everywhere and $R_{\rm d} = 
R_{\rm c} = 3$ (those cases are annotated with $\omega=1$ in
Table~\ref{table1}).

\subsection{Keplerian disc rotation profile}
\label{subsec:rdrceq}

The calculations presented here were performed using five values
of $C$ in the range  $3$--$120$ and using the Dirichlet
outer boundary condition.
 The maximum value of $R_{\rm ext}$ used
in these calculations decreases with $C$ (see
Table~\ref{table1}). As $C$ increases the resolution required to
resolve fully the magnetic field structure around the corotation
radius increases. We have therefore restricted the external radius for
$C =30$ or larger  so that a reasonable resolution of the area around
corotation can be achieved. 

The maximum value of $|B_\varphi/B_z|$ at the
disc surface found in the calculations with a Keplerian disc rotation
profile is $\sim 100$  being much larger than
values  considered by previous authors. Consequently we study the
effect of  increasing $|B_\varphi/B_z|$ on the force--free
magnetospheric field for a large region of  parameter space.

  Below we present results obtained with $C
= 3$ corresponding to maximum value of $|B_\varphi/B_z| =
2.56$--$2.97$ for $R_{\rm ext} = 11$--$80$. When $C=3$ we were able to
achieve the highest resolution and the largest range of $R_{\rm ext}.$

\subsubsection{Results with $C = 3$.}
\label{subsubsec:c3}

All the calculations with $C=3$ were performed with $Nr = 200$ and $Nm
= 80$ therefore  cases with smaller $R_{\rm ext} $ 
have better resolution.
\bfig
\noindent
\bmin{.48}
\centering\epsfig{file=./fig1a.ps,height=75mm,width=\linewidth,angle=270}
\emin \hfill
\bmin{.48}
\centering\epsfig{file=./fig1b.ps,height=75mm,width=\linewidth,angle=270}
\emin
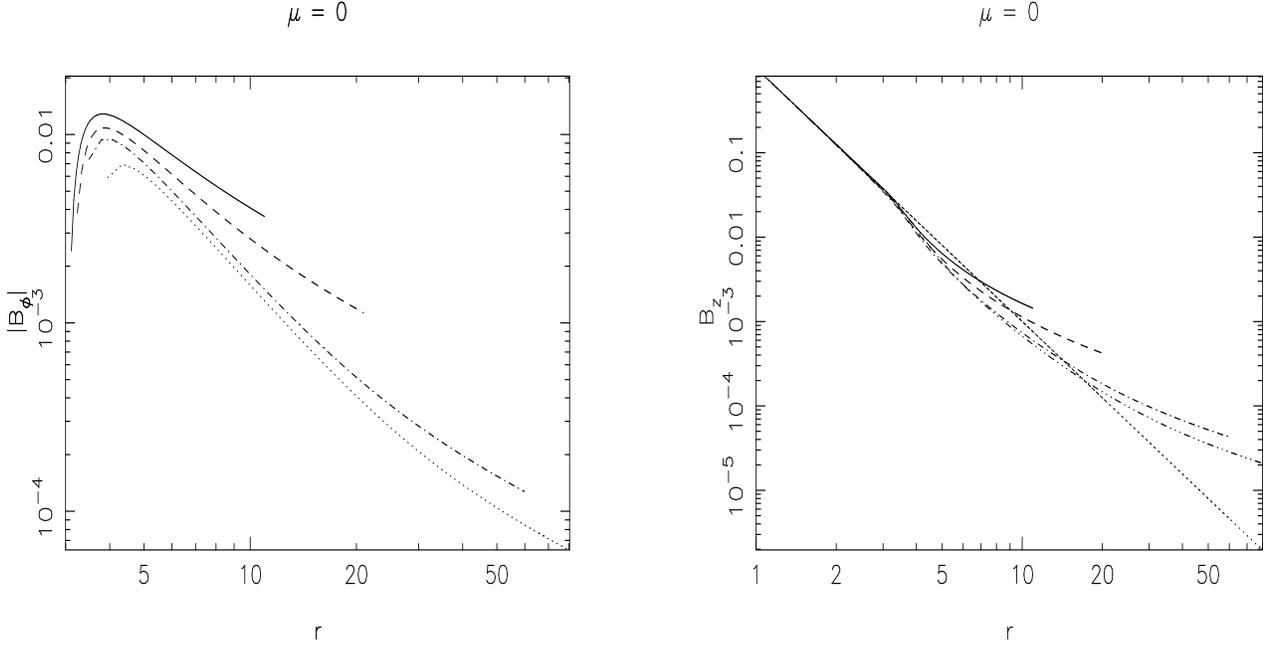
\caption{Results with $C=3$. {\it Left Panel:} Radial profile of
$|B_\varphi|$ at the disc surface with $R_{\rm ext} = 11$ (solid
line), $21$ (dashed line), $60$ (dot-dashed line) and $80$ (dotted
line). {\it Right Panel:} Radial profile of $B_z(r,0)$ with $R_{\rm
ext} = 11$ (solid line), $21$ (dashed line), $60$ (dot-dashed line)
and $80$ (triple dot-dashed line). The initial dipole profile of
$B_z(r,0)$ is represented by the dotted line.}
\label{bphirbzc3}
\efig

In the left panel of Fig.~\ref{bphirbzc3} we
plot the equilibrium radial profiles of $|B_\varphi|$ at the disc
surface. Since these profiles are well resolved we can conclude  that
increasing the computational domain leads to smaller values of $|B_\varphi|$. 
This  follows from the dependence of $B_z$ on $R_{\rm ext}.$
In the right hand panel of  Fig.~\ref{bphirbzc3}  we plot the radial
profile of $B_z$ at the disc surface for 
the different values of $R_{\rm ext}$.

The poloidal field  is similar  to the stellar dipole field
inside corotation  where $B_\varphi = 0$ but it deviates from
that significantly at all other radii. Immediately outside
corotation the field becomes smaller than the stellar  dipole
field as a result of field line inflation.
 At  the largest radii the poloidal field eventually becomes larger than the dipole
field locally. The radius at which this transition occurs
moves to larger radii as $R_{\rm ext}$ increases
indicating the influence of the outer boundary condition.
 The structure of the magnetic field is
not radially self--similar.  
The departure from self--similarity in the inner part of the
disc is due to the transition from $B_\varphi=0$ inside the corotation
radius to $B_\varphi$ given by equation (\ref{BPHIBZ}) for $R>R_{\rm
c}$ while  the outer
part of the disc is influenced by the outer
boundary condition. 

\bfig
\noindent
\bmin{.48}
\centering\epsfig{file=./fig2a.ps,height=75mm,width=\linewidth,angle=270}
\emin \hfill
\bmin{.48}
\centering\epsfig{file=./fig2b.ps,height=75mm,width=\linewidth,angle=270}
\emin \hfill
\bmin{.48}
\centering\epsfig{file=./fig2c.ps,height=75mm,width=\linewidth,angle=270}
\emin \hfill
\bmin{.48}
\centering\epsfig{file=./fig2d.ps,height=75mm,width=\linewidth,angle=270}
\emin
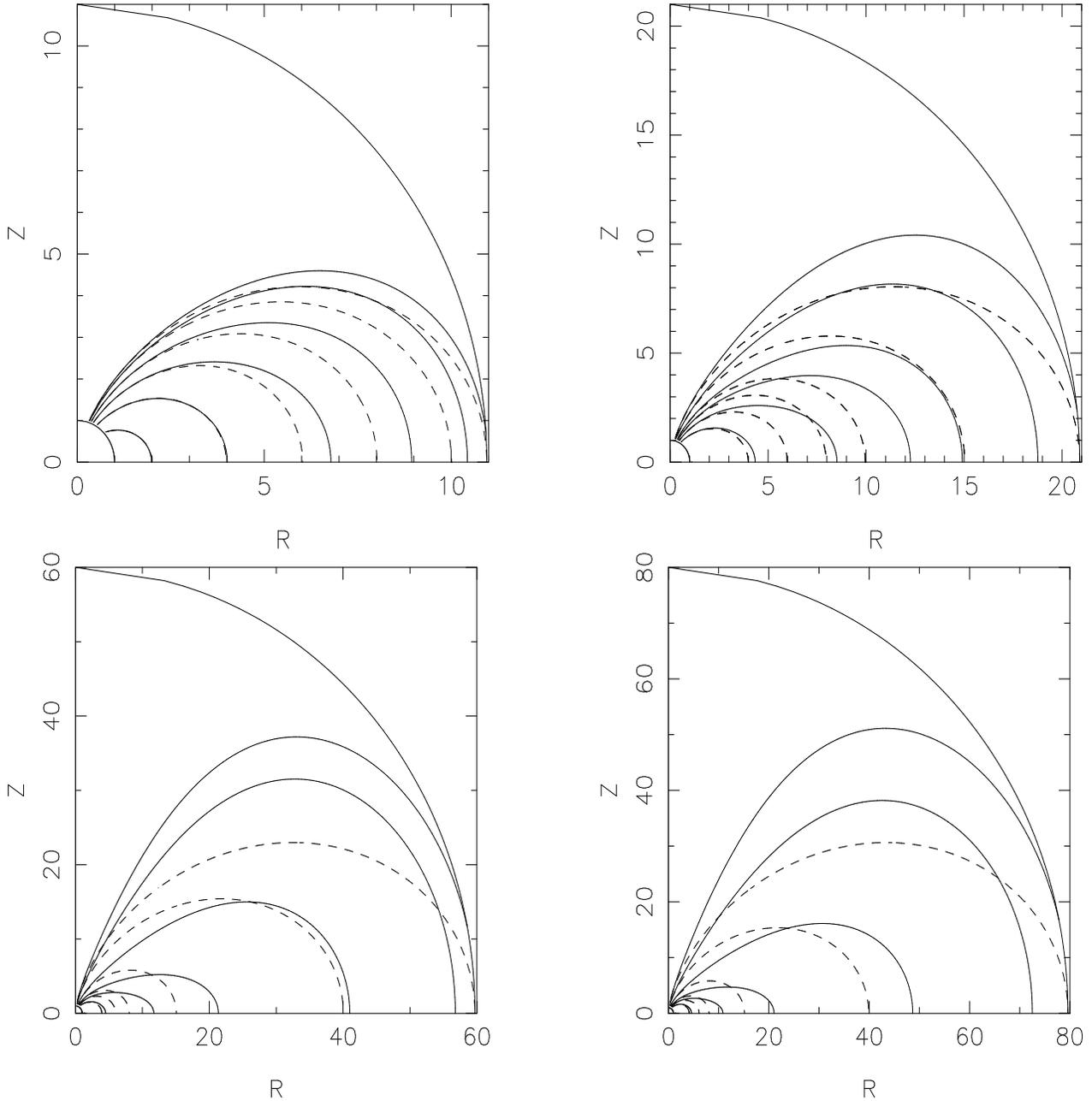
\caption{Isocontours of $\Psi_{\rm t}$ in the $R = r (1-\mu^2)^{1/2},
z = r \mu$ plane. The contour levels correspond to initial values of
$\Psi_{\rm t}$ (dipole contours plotted with dashed lines) for 
specific values of $r$ (and $\mu = 0$). Those are: {\it Upper left
panel:} $r= 1,2,4,6,8,10,11$, {\it Upper right
panel:} $r= 1,4,6,8,10,15,21$,{\it Lower left
panel:} $r= 1,4,6,8,15,40,60$, {\it Lower right
panel:} $r= 1,4,6,8,15,40,80$.}
\label{psicntc3}
\efig

Field line expansion as 
seen in Fig. (\ref{psicntc3}) for all $R_{\rm ext}$ used
is a consequence of  field line
twisting  together  with the assumption of a  force--free equilibrium
(see our discussion in Section~\ref{subsec:Crange}). The inflation of 
field lines increases with increasing $C$ or field line footpoint shear until the
field lines become partially or fully open (\citeNP{aly85}; 
\citeNP{wolfsonl92}; \citeNP{nnl92}; \citeNP{lb94}; \citeNP{aly95};
\citeNP{wolfson95}). In the fully open state line   twisting
 disappears and $B_\varphi = 0.$ 
\bfig
\centerline{\epsfig{file=./fig3.ps,height=100mm,angle=270}}
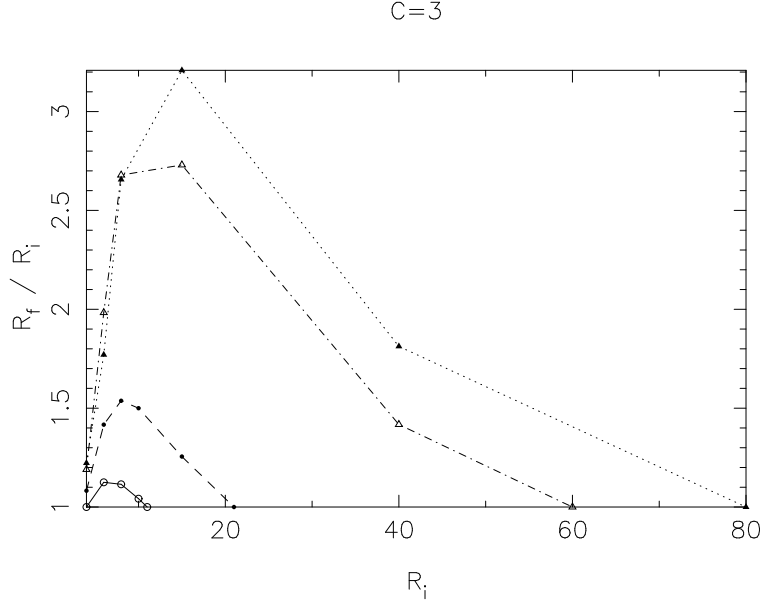
\caption{$R_{\rm f}/R_{\rm i}$ as a function of $R_{\rm i}$, where
$R_{\rm f}$ is the force--free and $R_{\rm i}$ the dipole footpoint
radii for $C=3$ with $R_{\rm ext} = 11$ (open circles),
$21$ (filled circles), $60$ (open
triangles) and $80$ (filled triangles).}
\label{rmaxc3}
\efig

In Fig.~\ref{rmaxc3} we plot the ratio of the final $R_{\rm
f}$ to the initial or dipole  $R_{\rm i}$ footpoint radius as a function of
$R_{\rm i}$ for a group of field lines. The values of $R_{\rm i}$
chosen are the same as in Fig.~\ref{psicntc3} with
$R_{\rm i} \ge 4$. Different curves correspond to different values of
$R_{\rm ext}$. The ratio $R_{\rm f}/R_{\rm i}$
increases linearly with $R_{\rm i}$ until a turnover radius is reached
where the outer boundary  starts affecting the rate of
inflation. For $R_{\rm i}$ larger than the turnover radius $R_{\rm
f}/R_{\rm i}$ decreases with $R_{\rm i}$. For the cases with $R_{\rm
ext} = 60$ and $80$ the values of $R_{\rm f}/R_{\rm i}$ are similar
for $R_{\rm i} \le 10$. We
therefore conclude that the maximum inflation has been reached for
the innermost disc region in these cases. The maximum overall value of
$R_{\rm f}/R_{\rm i}$ is $3.2$ and it corresponds to $|B_\varphi/B_z|$
of $2.73$. The maximum value of $R_{\rm f}/R_{\rm i}$ quoted by
\citeN{bardou} is 1.9 but this is an underestimate because of
resolution problems (A.Bardou, private communication).
Due to the effect of the outer boundary in our calculations it is difficult to
predict how inflated the configuration would become
 for $R_{ext} \rightarrow \infty.$
However, we  expect the field lines will remain closed
within a radius of approximately $3 R_{\rm c}$ for $C = 3$.

\subsubsection{Results for larger values of $C$}
\label{subsubsec:allc}
\bfig
\noindent
\bmin{.48}
\centering\epsfig{file=./fig4a.ps,height=75mm,width=\linewidth,angle=270}
\emin \hfill
\bmin{.48}
\centering\epsfig{file=./fig4b.ps,height=75mm,width=\linewidth,angle=270}
\emin
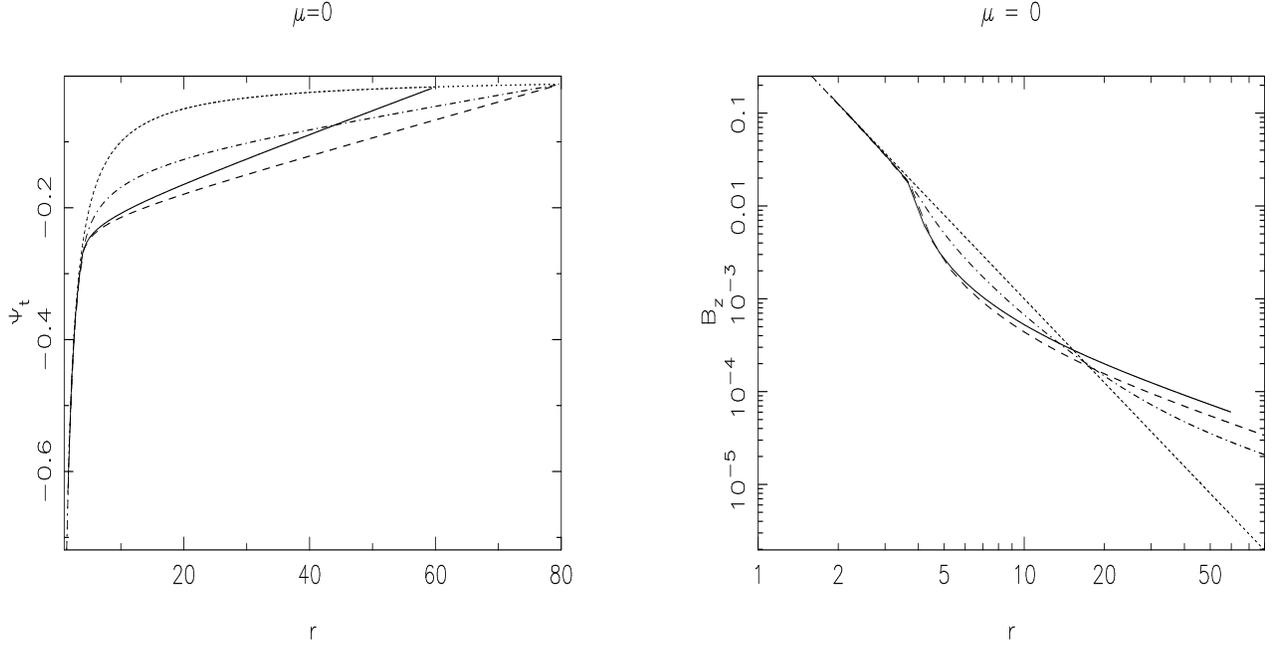
\caption{{\it Left Panel:} Radial profile of $\Psi_{\rm t}(r,0)$ 
for $C=10$ with $R_{\rm ext}=60$ (solid line) and $80$
(dashed line) and for $C=3$ with $R_{\rm ext}=80$ (dot-dashed
line). {\it Right Panel:} Radial profile of $B_z(r,0)$ for the same parameters as in the left panel. In both panels the initial dipole
profiles are represented by the dotted line.}
\label{totbzc3t10}
\efig

We performed calculations  for $C=10$  with $R_{\rm ext}=60$
and $80.$ Both these have $Nr=100$ and $Nm=50$ so the resolution is 
reduced with respect to the $C=3$ case. In Fig.~\ref{totbzc3t10} we
plot the radial profiles of $\Psi_{\rm t}(r,0)$ and $B_z(r,0)$ for $C=10$ and also for $C=3$ and $R_{\rm
ext}=80$. For $R_{\rm ext}=80$ we find that $\Psi_{\rm t}(r,0)$ is smaller for $C=10$ than for $C=3$ everywhere
in the disc as expected. For the smaller $R_{\rm ext}$ value the
effect of the outer boundary is stronger as it was found also for
$C=3$. We find that $B_z$ decreases with $C$ for $R < 15$. For larger radii
the trend is reversed because of the effect of the outer boundary. 
\bfig
\noindent
\bmin{.48}
\centering\epsfig{file=./fig5a.ps,height=75mm,width=\linewidth,angle=270}
\emin \hfill
\bmin{.48}
\centering\epsfig{file=./fig5b.ps,height=75mm,width=\linewidth,angle=270}
\emin
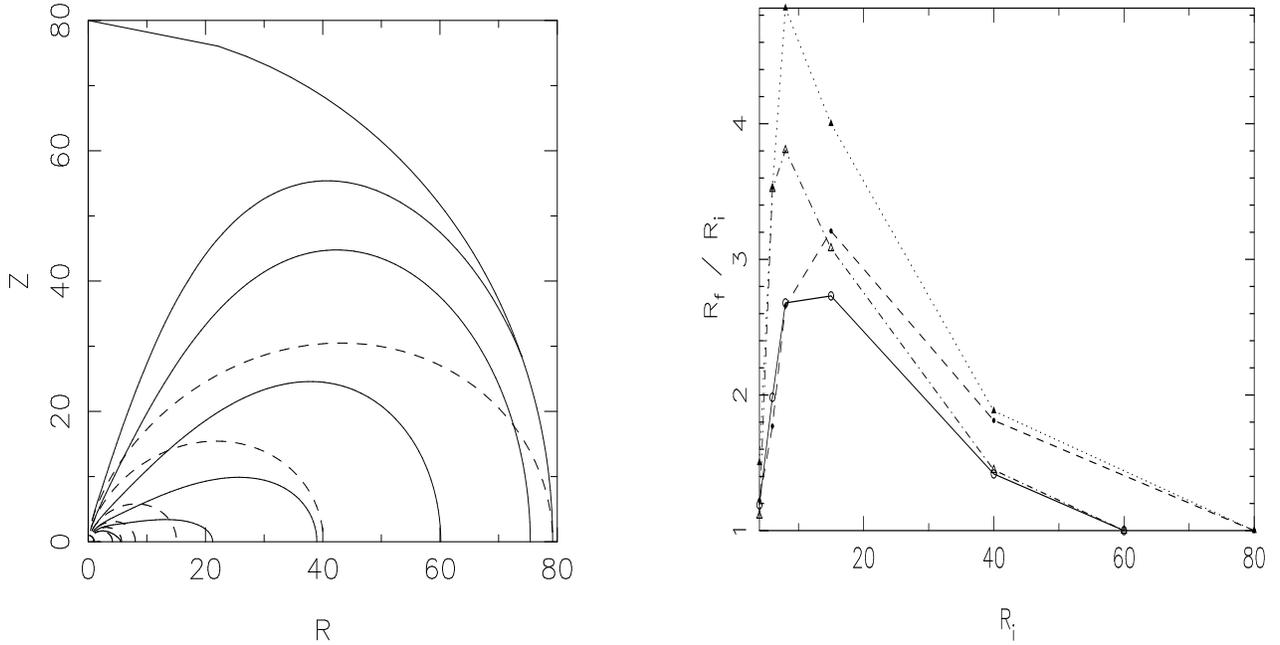
\caption{{\it Left panel:} Isocontours of $\Psi_{\rm t}$ on the $R,z$
plane plotted for $C=10$ with $R_{\rm ext} = 80$. The initial dipole
contours are plotted with dashed lines. The radii chosen are given in
Fig.~\ref{psicntc3}. {\it Right Panel:} $R_{\rm f}/R_{\rm i}$ as a
function of $R_{\rm i}$ for $C=3$ with $R_{\rm ext} = 60$ (open circles) and 
$80$ (filled circles) and for $C=10$ with $R_{\rm ext} = 60$ (open
triangles) and $80$ (filled triangles).}
\label{psicntc10r60t80} 
\efig

In the left panel of Fig.~\ref{psicntc10r60t80} we
plot  dipole and  equilibrium poloidal field lines for $C=10$ with
$R_{\rm ext} = 80$. Field lines 
interior to  $2R_{ext}/3$
are characterised by the largest flattening  at small $\theta$ while the
outer field lines are more spherical due to the 
 effects of the outer boundary. 
A  value of $R_{\rm ext}$
 exceeding the outer disc radius by a factor $10$--$100$ might be necessary 
in order to remove the effects of the outer boundary 
 on the calculation of field line inflation (see
\citeNP{roumeliotis94}). 

In the right panel of Fig.~\ref{psicntc10r60t80} we plot $R_{\rm
 f}/R_{\rm i}$ as function of $R_{\rm i}$ for $C=3$ and $C=10$, each
 with $R_{\rm ext} =60$ and $80$. For $R_{\rm i} \le 8$ the ratio
$R_{\rm f}/R_{\rm i}$ is $1.75$ times larger for $C=10$ than
for $C=3$ showing greater inflation. But 
for larger values of $R_{\rm i}$ this factor drops
to $1.25$. The larger influence of the outer boundary  on
the equilibrium for $C=10$ is manifested in the smaller turnover
radius observed in this case. It is not clear if the
ratio of final to initial footpoint radius found for $C=10$ at $R_{\rm
i} = 6$ would continue to increase linearly with $R_{\rm i}$ 
if the outer boundary effect was not
present. In the \citeN{bardou} results $R_{\rm f}/R_{\rm i}$ increases
linearly with $R_{\rm i}$ for $R \gg R_{\rm c}.$
 In our case $|B_\varphi/B_z|$ tends to a constant
for $R \gg R_{\rm c}$ and  $R_{\rm f}/R_{\rm i}$  could behave
similarly.
\bfig
\centerline{\epsfig{file=./fig6.ps,height=100mm,angle=270}}
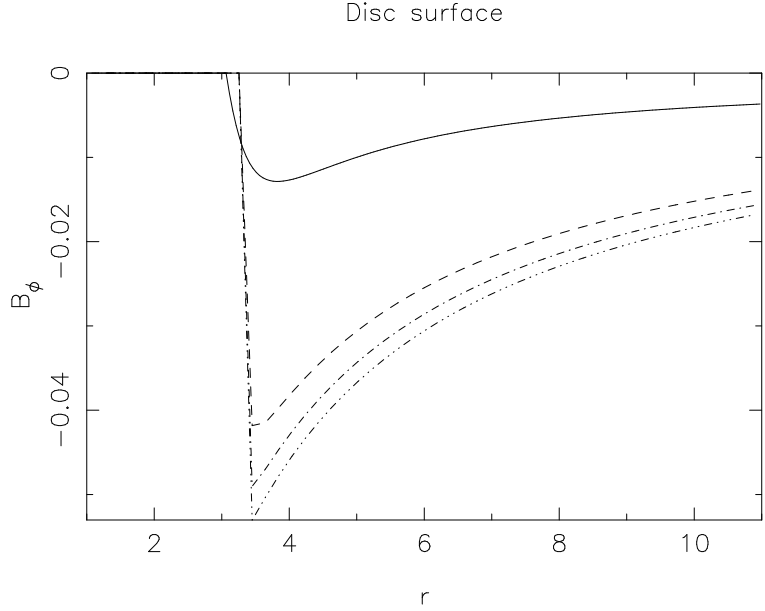
\caption{$B_\varphi$ as a function of radius for $C=3$ (solid
line), $C=30$ (dashed line), $C=60$ (dot-dashed line) and $C=120$
(triple dot-dashed line).}
\label{bphic3t120}
\efig

We now  discuss  results obtained for $C=30$--$120$. These
calculations were performed with $Nr=50$ and $Nr=30$ and $R_{\rm
ext}=11$ except for one case with $C=30$ and $R_{\rm
ext}=21$ which had $Nr=120$ and $Nm=50$. 
The disc surface equilibrium profiles of $B_\varphi$  
 for $C = 3$--$120$ and $R_{\rm ext} = 11$ 
are compared  in Fig.~\ref{bphic3t120}. 
Note that the magnitude of $B_{\varphi}$   varies only by
a factor of $3$ as $C$ varies between $3$ and $120.$

The effect of the growing
$|B_\varphi|$ 
 on the  stream function  for $C \ge
30$ and for $R_{\rm ext} = 11$  
 is to  make it  larger than the dipole 
value, more so around $R = R_{\rm c},$ leading to an
 enhancement
of $B_z(r,0)$ there. The disc surface
 radial profiles of $B_z$ for $C=3$--$120$  are compared in the left
panel of Fig.~\ref{bzlog_c30t120}. 
Because the toroidal magnetic pressure increases sharply 
 just outside $R=R_{\rm c}$ it is expected that 
   a larger poloidal field just interior
to $R_c$  is required for equilibrium.  Thus
field line deflation results. This  occurs in this model
 because  the field
lines are not sheared  inside corotation. For $R > R_{\rm c}$
$B_z(r,0)$ becomes 
smaller than the dipole  value before 
 increasing  close to the outer boundary.  
  Due to the increased $|B_\varphi/B_z|$ at the disc surface, 
$B_z$ decreases with $C$ outside corotation as   
expected. 

To study the  dependence 
on the outer  boundary radius for large $C$, we performed a calculation
with $C=30$ and $R_{\rm ext} = 21$ and compared the result to the case
where $R_{\rm ext} = 11$. The radial resolution is approximately the same
for these cases (see Table~\ref{table1}). The corresponding 
 radial profiles of $B_z(r,0)$ are plotted in the right panel of
figure~\ref{bzlog_c30t120}. 
\bfig
\noindent
\bmin{.48}
\centering\epsfig{file=./fig7a.ps,height=75mm,width=\linewidth,angle=270}
\emin \hfill
\bmin{.48}
\centering\epsfig{file=./fig7b.ps,height=75mm,width=\linewidth,angle=270}
\emin
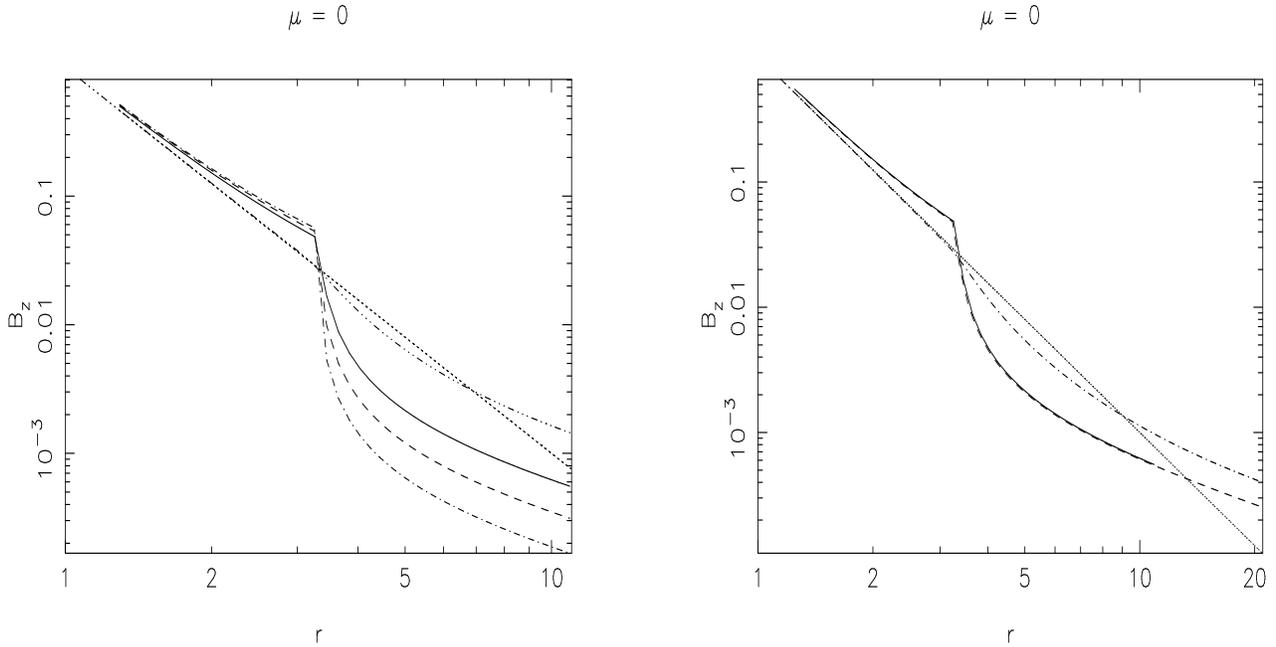
\caption{Radial profile of $B_z(r,0)$. {\it
Left panel:} $C=30$ (solid line), $C=60$ (dashed line), $C=120$
(dot-dashed line) and $C=3$ (triple dot-dashed line). All cases have
$R_{\rm ext} = 11$.{\it Right panel:} $C=30$ with $R_{\rm ext}=11$ (solid
line) and with $R_{\rm ext}=21$ (dashed line) and $C=3$ (dot-dashed
line). The initial dipole profile is represented by
the dotted line.}
\label{bzlog_c30t120}
\efig
When $R_{\rm ext}$ is increased $B_z(r,0)$ values are unmodified
inside $R = 10$ so the deflation of the field lines there  is
independent of the outer disc radius. The exterior
point at which $B_z(r,0)$ exceeds the dipole value  moves
to a larger radius for $C=30$ as compared to that when $C=3$. There is
also a change to the radial dependence of $B_z(r,0)$ which becomes
flatter at large $R$ for $C \ge 30$.  

\bfig
\noindent
\bmin{.48}
\centering\epsfig{file=./fig8a.ps,height=75mm,width=\linewidth,angle=270}
\emin \hfill
\bmin{.48}
\centering\epsfig{file=./fig8b.ps,height=75mm,width=\linewidth,angle=270}
\emin
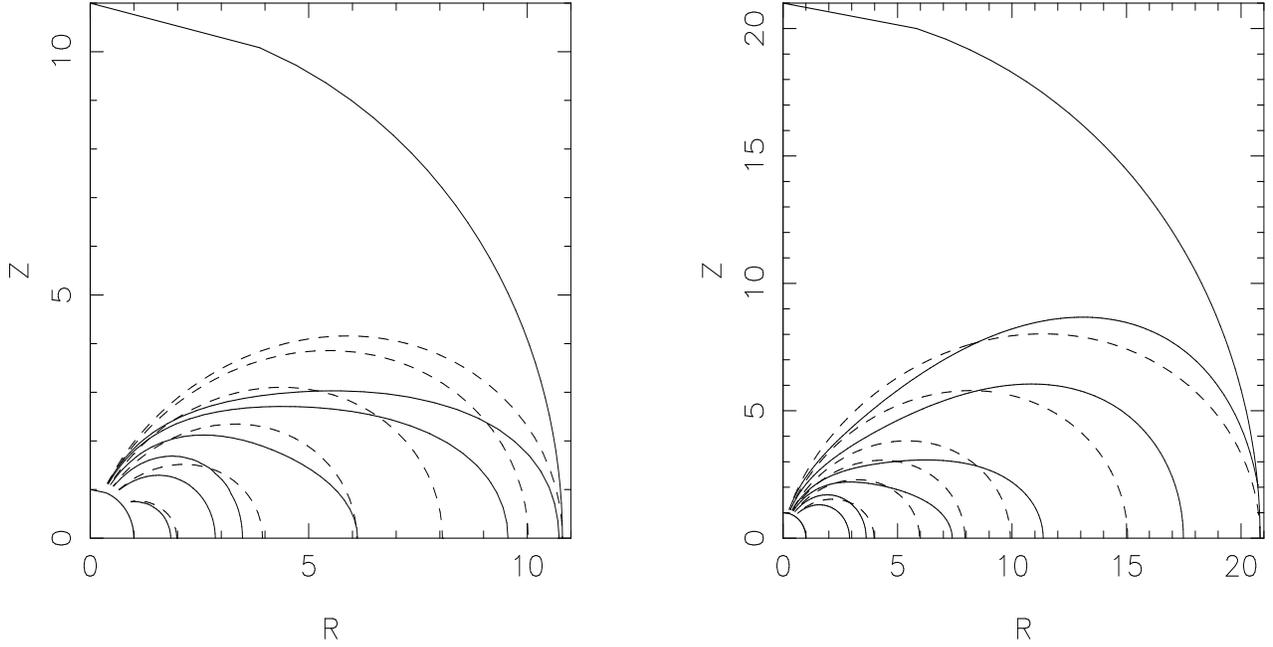
\caption{Isocontours of $\Psi_{\rm t}$ on the $R,z$ plane plotted
for $C=30$ with $R_{\rm ext} = 11$ (left panel) and with $R_{\rm ext}
= 21$ (right panel). The initial dipole contours are plotted with dashed lines
and the radii chosen are those already given in Fig.~\ref{psicntc3}.}
\label{psicntc30r11t21} 
\efig
In Fig.~\ref{psicntc30r11t21} we plot the
magnetic field lines for
 $C=30$ with $R_{\rm ext}
=11$ (left panel) and $R_{\rm ext} = 21$ (right panel). We see that
the deflation of field lines for $r < 10$ is  the same in
these cases.  At larger radii inflation rather than
deflation is observed for $R_{\rm ext} = 21$.
The field line inflation observed at $r>10$ for $C=30$ with $R_{\rm
ext}=21$ is smaller than that found for $C=3$ and the same  $R_{\rm
ext}$ (compare with Fig.~\ref{psicntc3}). This result is unexpected given the
larger $|B_\varphi/B_z|$ at the disc surface in the $C=30$ case. 
This is probably the
result of the increased influence of the outer boundary condition on
the equilibrium configuration for increasing $C$ values. 

In the next section we will discuss
results obtained using a sub--Keplerian form of $\Omega$ and which
results to a
smoother radial variation of $B_\varphi/B_z$ on the disc surface.

\subsection{Sub--Keplerian disc rotation profile}
\label{subsec:newom}

As the magnetic
field strength increases with decreasing disc radius and the magnetic
stresses start to  affect the radial and vertical disc equilibrium
the disc is expected to  ultimately corotate with the
star. Interior to the corotation radius the disc will then
be sub--Keplerian. In the present study we investigate  
the effect of  sub--Keplerian  rotation  
on $B_\varphi/B_z$ and the equilibrium poloidal 
magnetic field. We modify $\Omega$ 
so that it becomes sub--Keplerian for $R \sim R_{\rm c}.$
The form of $\Omega$ that we will use is that of \citeN{campbell87}
which is given, for $R > R_{\rm d}$, by:

\beq
\Omega = \Omega_{\rm K}(R_{\rm d}) \left\{\left(\frac{R_{\rm
d}}{R}\right)^{3/2} - (1-\omega) \exp{\left[-\frac{3}{2}(1-\omega)^{-1}
\left(\frac{R}{R_{\rm d}} -1\right)\right]}\right\}.
\label{OMNEW} \eeq
 
\noindent Here $\omega = \Omega
_{\star}/\Omega_{\rm K}(R_{\rm d})$. 
At $R = R_{\rm d}$ we have $\Omega =
\Omega_\star$ and ${\rm d}\Omega/{\rm d}R = 0$.
Interior to $R = R_{\rm d}$, the material
 is taken  to rotate  with $\Omega = \Omega_{\star}.$
We plot $\Omega$ as a function of $R$ in the left panel of
Fig.~\ref{om_new} for $R_{\rm c} = 3$ 
 and
$\omega = 0.875$ and $0.2$. The condition: $\Omega =\Omega_{\star}$ at
$R=R_{\rm c}$ determines $R_{\rm d}$ through the chosen value of $\omega$.
 
For the largest value of $\omega$, $R_{\rm d}$ almost coincides with
$R_{\rm c}$ and $\Omega$ remains Keplerian until very close to $R_{\rm c}$. For smaller
values of $\omega$ we have $R_{\rm d} < R_{\rm c}$ and in the case 
where $\omega = 0.2,$  $R_{\rm d}$  almost reaches  the stellar surface ($R =
1$). Here as in the previous sections we have $B_\varphi =0$ for $R <
R_{\rm d}$ but a  part of the differentially
rotating  disc may exist inside corotation 
and that can affect the equilibrium  magnetic field. 
The variation of $B_\varphi/B_z$ with $r$  at the disc surface
can be seen in the right panel of Fig.~\ref{om_new} where we took $C=30.$
\bfig
\noindent
\bmin{.48}
\centering\epsfig{file=./fig9a.ps,height=75mm,width=\linewidth,angle=270}
\emin \hfill
\bmin{.48}
\centering\epsfig{file=./fig9b.ps,height=75mm,width=\linewidth,angle=270}
\emin
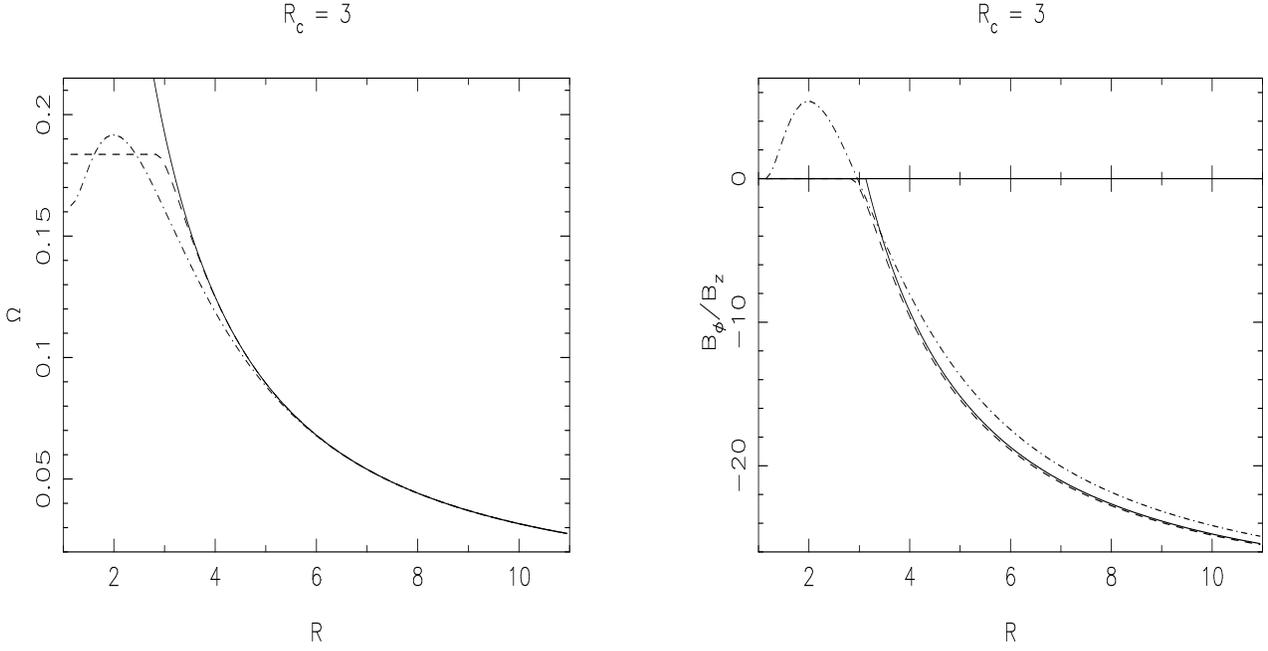
\caption {Radial profiles of $\Omega$ (left panel) given by
equation (\ref{OMNEW}) and of corresponding $B_\varphi/B_z$ (right
panel) using
$R_{\rm c} = 3$  with $\omega = 0.875$ (dashed line)
and $\omega = 0.2$ (dot-dashed line). The radial profiles of $\Omega_{\rm K}$ and corresponding
$B_\varphi /B_z$ are plotted with a solid line.}
\label{om_new}
\efig
For $\omega =
0.875$ the profile of $B_\varphi/B_z$ is virtually unchanged with
respect to the Keplerian case with $R_{\rm d} = R_{\rm c}.$ 
A larger effect on $B_\varphi/B_z$ occurs when 
 $\omega = 0.2. $ In  that case
there is a region inside corotation where $B_\varphi/B_z$ reverses
sign.  

\subsubsection{Results with $\omega = 0.875$ and $C=30$--$120$}
\label{subsubsec:om_0_875}

These calculations have $R_{\rm ext} = 60.$  
\bfig
\noindent
\bmin{.48}
\centering\epsfig{file=./fig10a.ps,height=75mm,width=\linewidth,angle=270}
\emin \hfill
\bmin{.48}
\centering\epsfig{file=./fig10b.ps,height=75mm,width=\linewidth,angle=270}
\emin
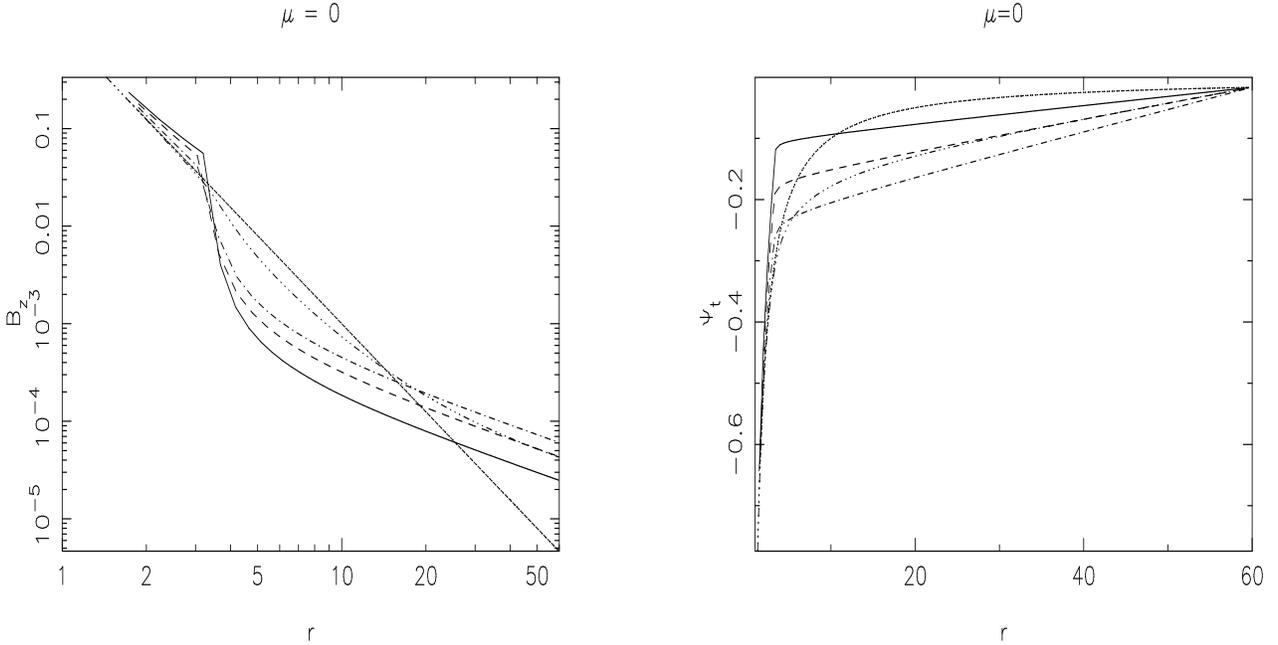
\caption{Radial profile of $B_z(r,0)$ (left panel) and of $\Psi_{\rm
t}(r,0)$ (right panel) for $\omega =0.875$ and $C=120$ (solid
line), $C=60$ (dashed line), $C=30$ (dot-dashed line) and $C=3$ for
$\Omega = \Omega_{\rm K}$ (triple dot-dashed line). The
initial dipole profiles are represented by a dotted line.}
\label{bzlogc3t120newom}
\efig
In the left panel of Fig.~\ref{bzlogc3t120newom} we plot $B_z(r,0)$
for $C=30$--$120$ with $\omega = 0.875$ and $R_{\rm ext} =60$ with
the corresponding $\Psi_{\rm t}(r,0)$ profiles shown in the right panel. 
We  also plot the case $C=3$  with $\Omega=\Omega_{\rm
K}$. Apart from boundary effects,
$|B_z|$ decreases  with $C$, the decrease 
being faster between $C=60$ and $120$ than between $C=30$ and $60$. 
These results reinforce our expectation: $B_z \rightarrow 0$ for 
$C \rightarrow \infty$.

As seen in the right panel of Fig.~\ref{bzlogc3t120newom} the disc
values of $\Psi_{\rm t}$ increase with $C$ between
$C=30$ and $120$. This is contrary to theoretical prediction and to the
trend observed between $C=3$ and $30$. As we will see in
Section~\ref{subsec:secbc} this result depends strongly on the
outer boundary condition. 

The profiles obtained in the non--Keplerian case with $\omega=0.875$
are smoother around
corotation than in the Keplerian case. This is due to the smoother
radial variation of $B_\varphi/B_z$. However, $B_z(r,0)$ is still larger
than the dipole value there and some deflation is observed interior to
$r=R_{\rm c}$. At larger radii field lines are inflated as we would
expect given the increase in $R_{\rm ext}$. 

\subsubsection{$\omega = 0.2$}
\label{subsubsec:om_0_2}
 
In this case $R_{\rm d}/R_{\rm c}=0.34$ and 
 an inversion   of the sign of $B_\varphi$  may significantly  affect
the variation of $B_z$ with radius. 
 We adopted $C=3$ with $R_{\rm
ext} = 21$ and  $C=30$ with $R_{\rm ext} =11$ with two different
resolutions (see Table~\ref{table1} for details). For $C=3$ the
difference between Keplerian and non--Keplerian cases is not significant. 

In the left panel of Fig.~\ref{bphibzc30of02newom} we
plot the  disc surface
profiles of $B_\varphi$ for $C=30$ and $\omega=0.2$ obtained  with two
different  resolutions and for $C=30$ with $\Omega=\Omega_{\rm K}$.
For $C=30$ the larger radial variation of
$|B_\varphi|$ in the non--Keplerian case has a more significant effect
on the equilibrium.
As the radial gradient of
$B^2_\varphi$ increases for small radii a small deflation of the
field lines is observed there. Eventually $B^2_\varphi$ decreases with
radius and inflation is observed for $R > 2$. The reverse was observed
in the Keplerian case where field lines were found to be deflated
everywhere in the disc (always for $R_{\rm ext} = 11$).

In the right panel of Fig.~\ref{bphibzc30of02newom} we plot the radial
profile of $B_z(r,0)$ for the same parameters as in the left panel.
We note that when $\omega=0.2$ the field deviates
more from the dipole form around corotation, 
becoming larger than the dipole value locally, than in the case with
$\Omega=\Omega_{\rm K}.$ This difference is mainly due to the larger
radial variation of $|B_\varphi|$ with radius in the non--Keplerian
case. The difference observed between the profiles obtained in the
high and low resolution cases is mostly due to the radial shift of the
gridpoint corresponding to $R=R_{\rm c}$. 
\bfig
\noindent
\bmin{.48}
\centering\epsfig{file=./fig11a.ps,height=75mm,width=\linewidth,angle=270}
\emin \hfill
\bmin{.48}
\centering\epsfig{file=./fig11b.ps,height=75mm,width=\linewidth,angle=270}
\emin
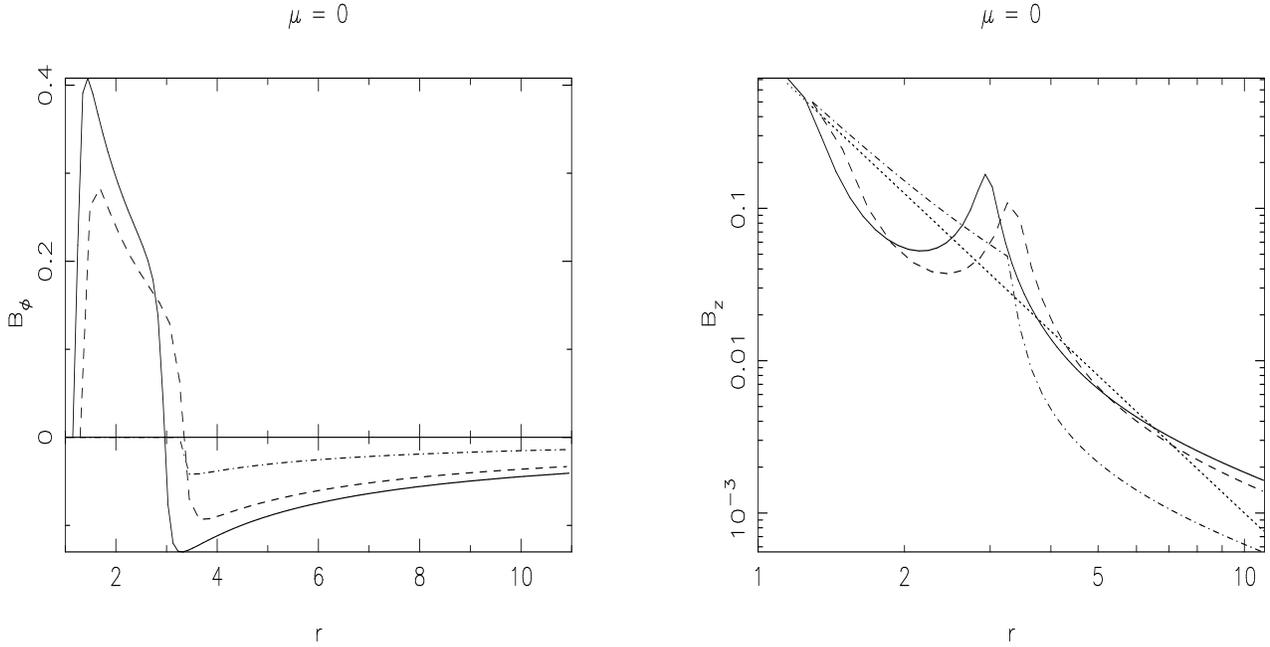
\caption{{\it Left Panel:} Radial profile of $B_\varphi$ at the disc
surface for $C=30$ with 
$R_{\rm ext}=11$ and for $\omega=0.2$ with $Nr=100$ (solid line) and 
with $Nr=50$ (dashed line) and for $\Omega=\Omega_{\rm K}$ (dot-dashed
line). {\it Right Panel:} Radial profile of $B_z(r,0)$ for the same
parameters as in the left panel. The initial dipole profile is represented by
the dotted line.}
\label{bphibzc30of02newom}
\efig

We conclude that the deflation observed in previous
calculations with a Keplerian rotation profile and with $\omega=0.875$
is related to the restricted twisting of field lines inside the corotation
radius. In the case with $\omega = 0.2$ we have $B_\varphi \ne 0$
almost everywhere in the disc. Consequently field lines are much more
inflated in this case. 
 
\subsection{A different outer boundary condition}
\label{subsec:secbc}

To study the effect of the outer boundary condition on our results we
have also performed calculations with the Neumann condition:
${\upartial \Psi \over \upartial r} =0$ and for $C=3,10,60$ and
$120$.  
The  parameters used in these
calculations are given in Table~\ref{table1}. All cases have
$\omega=0.875$ except when $C=3$ where a Keplerian rotation profile
was used. The new boundary condition makes the disc field
normal to the outer boundary. Thus we expect some of the field lines
to be open at the equilibrium, at least at large disc radii far
from the disc midplane. 

Indeed the results show much larger inflation and opening of the field
lines in this case than in the cases presented in previous sections
where the Dirichlet  condition was used. Also almost no deflation is
observed inside corotation showing the global effect of the outer
boundary condition. As seen in
Fig.~\ref{psitbz_nbc} both $\Psi_{\rm t}(r,0)$ and $B_z(r,0)$
decrease with $C$ monotonically. For $C=120$ the equilibrium seems to
have reached a limiting configuration with almost all field lines open. 
\bfig	
\noindent	
\bmin{.48}	
\centering\epsfig{file=./fig12a.ps,height=75mm,width=\linewidth,angle=270}
\emin \hfill
\bmin{.48}
\centering\epsfig{file=./fig12b.ps,height=75mm,width=\linewidth,angle=270}
\emin
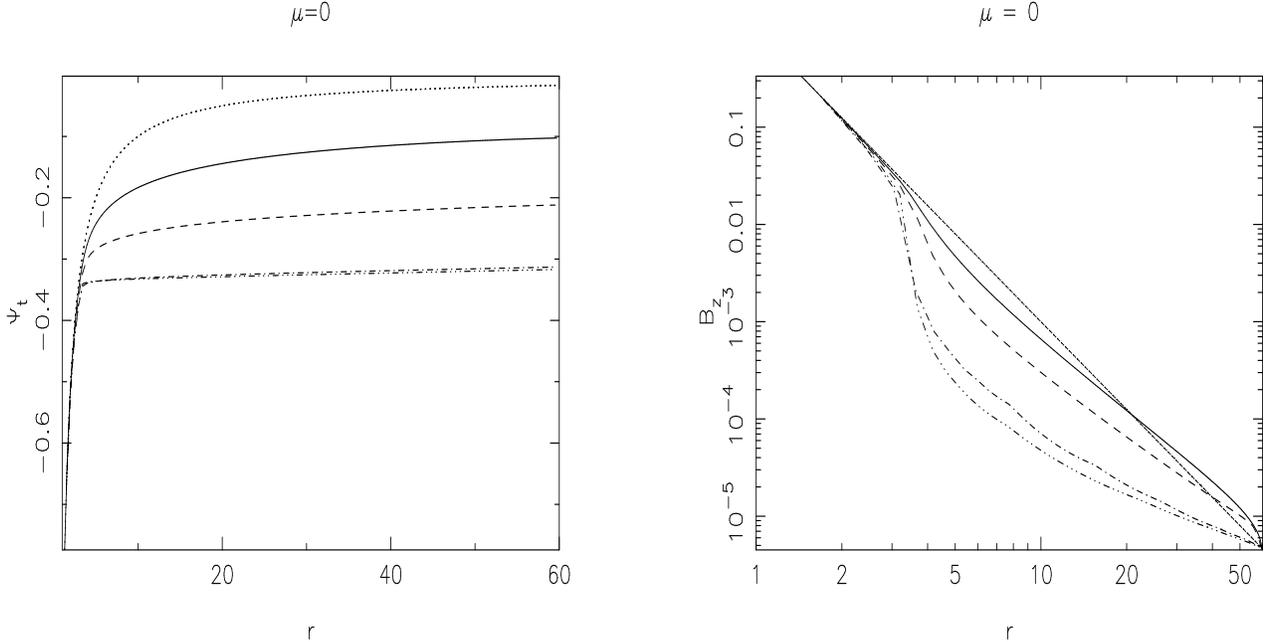
\caption{$\Psi_{\rm t}(r,0)$ (left panel) and $B_z(r,0)$ (right panel)
for $C=3$ (solid line), $10$ (dashed line), $60$ (dot-dashed line) and
$120$ (triple dot-dashed line). Note that the $\Psi_{\rm t}(r,0)$
profiles for the last two cases approximately coincide. The initial dipole profiles are represented by
the dotted line.}
\label{psitbz_nbc}
\efig

Comparing the disc radial profiles of $\Psi_{\rm t}(r,0)$ depicted here with 
those in Fig.~\ref{bzlogc3t120newom} we note that $\Psi_{\rm t}(r,0)$
decreases considerably when the second boundary condition is
used. Also $\Psi_{\rm t}(r,0)$ becomes approximately
independent of $r$ everywhere outside the corotation area which is the
limiting behaviour that we expect for large values of $C$. Comparing
the disc radial profiles of $B_z(r,0)$ between
figures~\ref{bzlogc3t120newom} and~\ref{psitbz_nbc} we note that for $C=3$
the magnetic field is similar for the two boundary conditions for $r
\le 10$. Further out $B_z(r,0)$ is smaller for the second boundary
condition. This difference increases with $C$. We also note that 
$B_z(r,0)$ has a stronger dependence on $r$ when the second 
boundary condition is used.

In the first three panels of Fig.~\ref{psicntnbc} we plot the magnetic
field lines for $C=3,10$ and $60$ respectively, using the same initial
(dipole) field lines as in Fig.~\ref{psicntc3} (with $R_{\rm ext} =
60$). For $C=3$ the level of inflation is comparable  for 
the two boundary conditions for $r \la 10$. For
larger initial footpoint radii field lines become much more inflated
for the second outer boundary condition and eventually they open up
for $r \ge 15$. As $C$ increases more field lines become open as seen
by comparing the cases with $C=10$ and $60$. The case with $C=120$ is virtually
identical to the $C=60$ case. Some field lines remain closed even when
$C=60$ as  can be seen by plotting more contour levels (see lower
right panel of Fig.~\ref{psicntnbc}). The closed field lines 
correspond to initial radii which are very close to the inner disc
radius. Consequently $ff'$ is non-zero for an increasingly smaller range of
$\Psi_{\rm t}$ values as $C$ increases. Calculations for
larger values of $C$ could only be performed accurately using  
considerably larger numerical resolution. However, our present calculations
have already shown that field lines at equilibrium
will be approximately open  for $|B_\varphi/B_z| \sim 10^2$
if a Neumman outer boundary condition is used. 
\bfig
\noindent
\bmin{.48}
\centering\epsfig{file=./fig13a.ps,height=75mm,width=\linewidth,angle=270}
\emin \hfill
\bmin{.48}
\centering\epsfig{file=./fig13b.ps,height=75mm,width=\linewidth,angle=270}
\emin \hfill
\bmin{.48}
\centering\epsfig{file=./fig13c.ps,height=75mm,width=\linewidth,angle=270}
\emin \hfill
\bmin{.48}
\centering\epsfig{file=./fig13d.ps,height=75mm,width=\linewidth,angle=270}
\emin
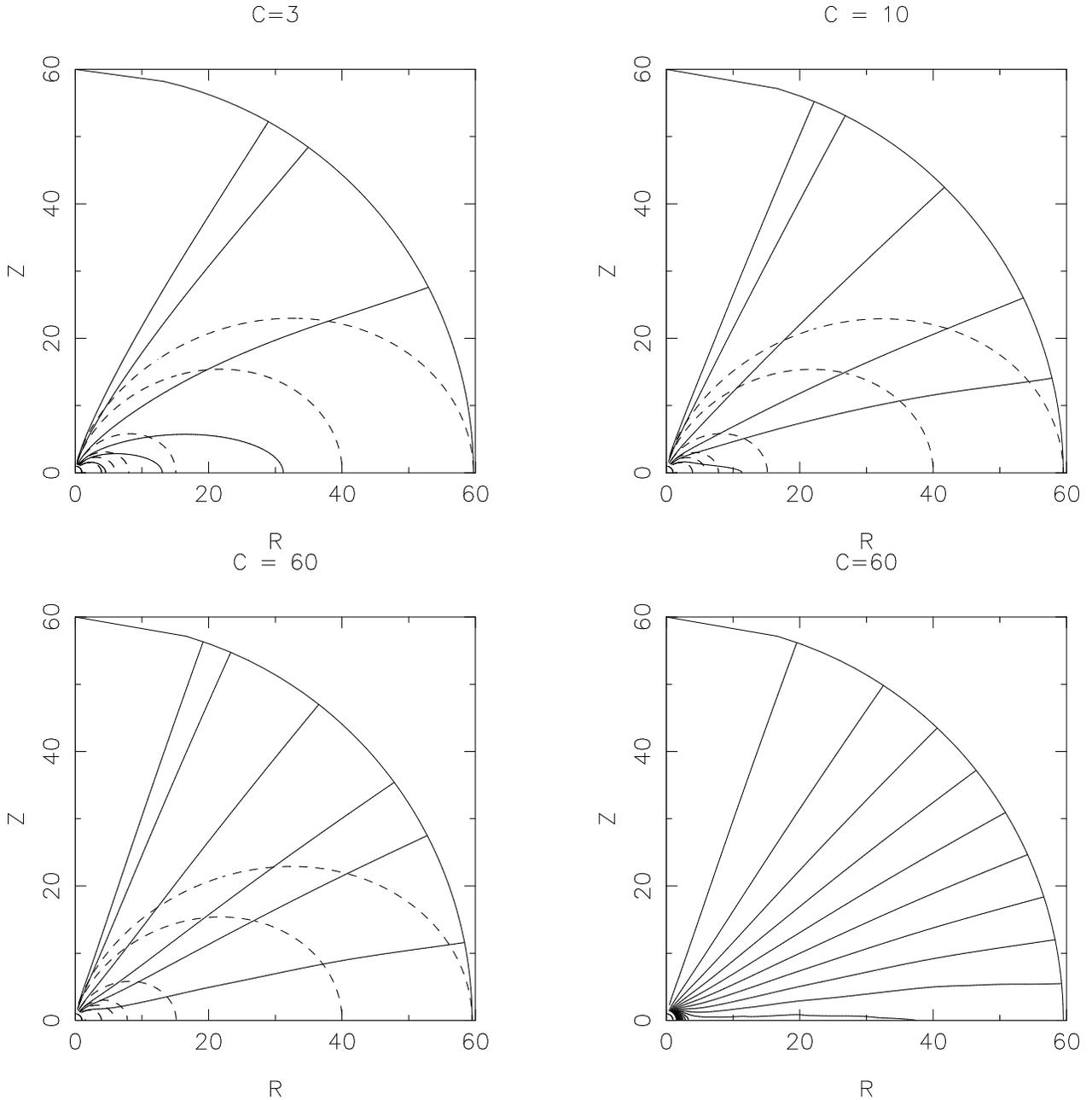
\caption{{\it Upper left to lower left panels:} Isocontours of $\Psi_{\rm t}$ on the $R,z$ plane plotted
for $C=3,10$ and $60$ using the initial footpoint radii given in
Fig.~\ref{psicntc3}. {\it Lower right panel:} Isocontours of
$\Psi_{\rm t}$ for $C=60$ using $20$ equally spaced contour levels.}
\label{psicntnbc}
\efig

The ratio $|B_\varphi|/B_{\rm p}$ is found to be less than unity for
$\mu < 0.1$. Therefore $B_\varphi \approx 0$ in most of the corona as 
expected  on open field lines. In the limit of a large
twist the disc field acts as to expel the
poloidal magnetic field from the disc and consequently to annihilate
the toroidal magnetic field, as long as the disc field is permitted
to penetrate the outer boundary. In the limit of large $C$ the
poloidal field resembles  the solution $Z_1$
given by \citeN{low86a}. This corresponds to a perfectly
conducting disc with central hole immersed in a central dipole field
where some flux escapes to infinity such that the field is non singular
at the disc inner edge.

\subsection{Modification of the spin--down and spin--up torques}
\label{subsec:torque}

For the force--free
equilibria calculated above, the  poloidal
magnetic field  differs significantly  from the original stellar dipole
form.  As a consequence, the local magnetic
torque acting on the disc
 may also  differ significantly  from what one obtains
assuming $B_z \sim B_{\rm dp}$.
Since this torque  regulates  the spin evolution of the central star
it is important to examine the possible  effect of our results on
the  calculation of the total torque experienced by the star. 

The total magnetic torque $N$
can be split into two components of opposite sign.
 The first
arises inside corotation where angular momentum
is transfered from the disc to spin up the star.  The  second 
arises outside corotation where angular momentum
is  transferred to the disc. In order for  stellar accretion to occur
 we require $R_{\rm d} < R_{\rm c}$. In  one  set of
 calculations we assumed $R_{\rm d} = R_{\rm c}$ and adopted
Keplerian rotation. In that case
the spin--up torque would be zero. In reality we expect $R_{\rm d}/R_{\rm c}
 \approx
0.91$--$0.97$ (i.e. \citeNP{wang95}) giving rise to a relatively small
spin--up torque. This torque is probably small compared to that
arising from direct accretion onto the star.

We now  calculate the total torques for the case where
the disc is assumed to be Keplerian and truncated at the corotation radius.

\subsubsection{The Keplerian case}
\label{subsubsec:r0rceq}

The total magnetic torque {\it on the star} is in general given by
(e.g. \citeNP{ghoshl79b}):

\beq
N = - \int^{R_{\rm ext}}_{R_{\rm d}} B_\varphi B_z R^2 {\rm d}R.
\label{TORQUE}
\eeq

\noindent We adopt dimensionless
parameters such that the torque  is given in units of $N_\star \equiv
B^2_\star r^3_\star$. Of course all field values are calculated on the disc
surface ($\mu=0$). For the cases presented here the first boundary
condition has been used. When $R_{\rm d} = R_{\rm c},$ as is considered
here, $N$ is positive definite and  only acts to spin down the star.

The total torque $N$ can be calculated analytically for a dipole
 field with $B_\varphi$ given by equation
(\ref{BPHIBZ}) and  is given by:

\beq
N^{\rm dp} = \frac{C}{3} \left[ \frac{1}{3} +
\frac{2}{3} \left(\frac{R_{\rm ext}}{R_{\rm c}}\right)^{-9/2} -
\left(\frac{R_{\rm ext}}{R_{\rm c}}\right)^{-3} \right] R^{-3}_{\rm
c}.
\label{NDOWN1_A}
\eeq

In the left hand panel of
Fig.~\ref{torquec3t120} we plot $N/C$ and $N^{\rm dp}/C$ as a function 
of $C$ for $R_{\rm ext}=11$. Although $N^{\rm dp}/C$ is constant
with $C$, $N/C$ decreases with increasing $C$ as the poloidal
magnetic field in the force--free equilibria decreases  with increasing
$C$. 
The  ratio of force--free to dipole torque $N/N^{\rm
dp}$ for $C \ge 30$ varies between $0.14$--$0.013$ which is significantly
smaller than the same ratio in the case with $C=3$. Although the
effect of the outer boundary condition is to suppress inflation in all
cases with large $C$ values and $R_{\rm ext}=11$, the effect of the
twisting of the field lines on the final 
equilibrium  is an increased reduction of the total torque 
with respect to its corresponding dipolar value.
\bfig
\noindent        
\bmin{.48}
\centering\epsfig{file=./fig14a.ps,height=75mm,width=\linewidth,angle=270}
\emin \hfill
\bmin{.48}
\centering\epsfig{file=./fig14b.ps,height=75mm,width=\linewidth,angle=270}
\emin
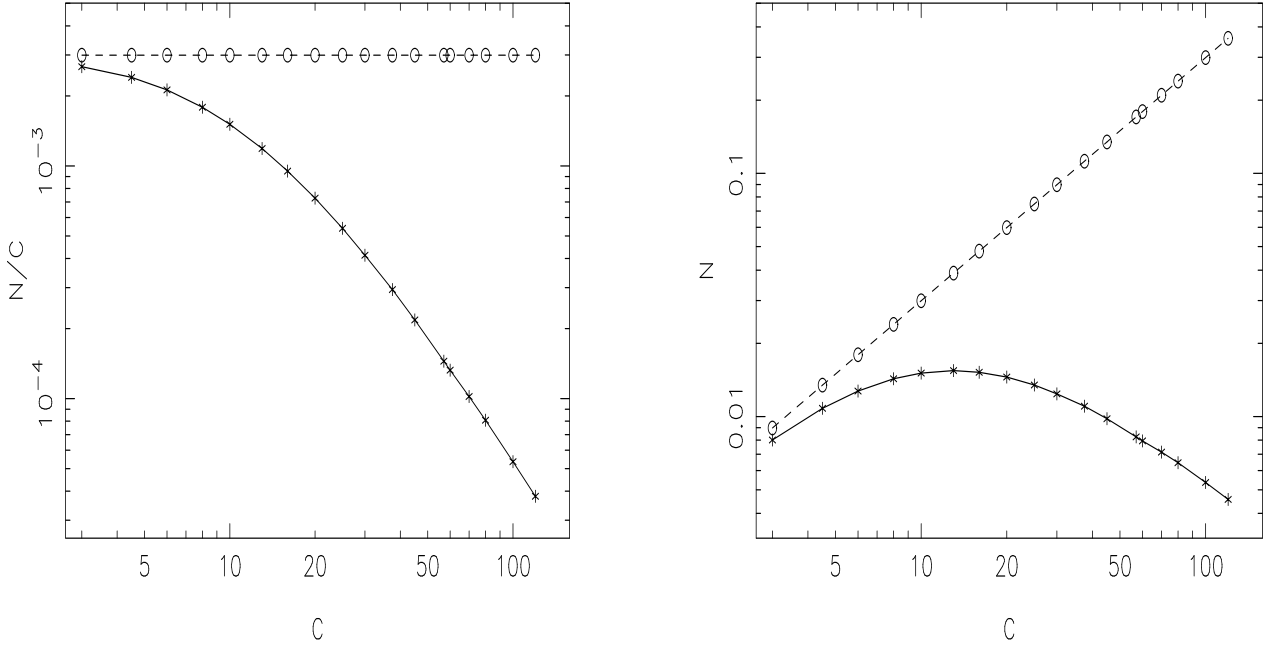
\caption{{\it Left Panel:} $N/C$ as function of $C$ for
$R_{\rm ext} = 11$. {\it Right Panel:} $N$ as function of $C$ for
$R_{\rm ext} = 11$. Force--free values are marked with stars
(solid line) and dipolar values are marked with circles (dashed
line). Many more models are employed here than those tabulated in
Table~\ref{table1}. The torque is seen to vary smoothly with $C.$} 
\label{torquec3t120}
\efig

In the right panel of
Fig.~\ref{torquec3t120} we plot $N$ and $N^{\rm dp}$ where we see that
the behaviour of $N$ as function of $C$ is very different (in fact
almost reversed) to that expected for a dipolar  poloidal field. From this
figure we see that $N$ increases slowly for $C \le 15$ and it
decreases for larger values of $C$. As the 
 values of $N$ for $C \ge 30$ are between $10$ and $100$ times smaller
than the dipolar values the increasing field line twisting  has
an important effect on the spin down of the star. The
magnetic breaking  timescales calculated are longer and
therefore magnetic breaking models might
have to be reevaluated.
However, we have not
taken into account here the disc
region inside corotation.  Also the results described here
for $C \ge 30$ have  $R_{\rm ext}=11$ which is relatively
small and therefore the effect of the outer boundary condition is 
important.

\subsubsection{The sub--Keplerian case}
\label{subsubsec:omeganK}

In the case where $\Omega \ne \Omega_{\rm K}$ the total magnetic
torque includes both spin--up and spin--down components. For the
case with $\omega = 0.875$ the spin--up component is of no
significance. 
In the following we will compare the total magnetic torques
 calculated using a sub--Keplerian
rotation profile with $\omega=0.875$ to the corresponding dipolar
values. 
Calculations with similar
parameters were performed with both outer boundary conditions.

In Fig.~\ref{torquesc3t120of} we plot $N/C$ and $N^{\rm dp}/C$ 
as a function of $C$. The values in the left panel of this figure
are from calculations performed with the first outer boundary
condition. Here, as for the Keplerian case, we find that $N/C$
decreases with $C$ and that the ratio $N/N^{\rm dp}$ ranges from
$0.56$ for $C=3$ to $0.05$ for $C=120$. The larger ratios here
are mainly a consequence of the use of a sub--Keplerian rotation
profile. However, For $C\ge 30$ we have again $N/C \propto 1/C$
approximately. Therefore, the trend for a smaller total torque is clear and
the difference between 
force--free and dipole cases is again significant for $C \gg 1$. For
$B_\varphi \sim B_z$ the difference is not as large although we have
to reserve final judgement until calculations with much larger values
of $R_{\rm ext}$ are performed.

For  calculations performed with the second outer boundary condition
we found significantly smaller values  
of $B_z(r,0)$  which decreases faster with $r$ 
than when the first boundary condition is used. 
The steeper decline of $B_z(r,0)$ with $r$ is more noticeable in the cases
with large $C$ values. For large
$C$ values we therefore expect much smaller values of $N/C$ to arise when the
second boundary condition is used. In the right panel of
Fig.~\ref{torquesc3t120of} we plot $N/C$ and $N^{\rm dp}/C$ as a
function of $C$ for the calculations presented in
Section~\ref{subsec:secbc}. The ratio $N/N^{\rm
dp}$ varies in the range $0.45$--$0.003$. Although this ratio is almost
unity for $C=3,$ as was  the case for the first boundary
condition, it becomes progressively smaller for larger $C$ values.
For large values of $C$ we have $N/C \sim 1/C^{3/2}$ when the second
boundary condition is used. 

In summary, when
 $|B_\varphi/B_z|$ on the disc surface is large 
the   magnetic torque acting on it 
will be  much
smaller than that estimated  assuming $B_z \sim B_{\rm dp},$
especially when  external conditions enable the opening of field lines.

\bfig
\noindent	
\bmin{.48}
\centering\epsfig{file=./fig15a.ps,height=75mm,width=\linewidth,angle=270}
\emin \hfill
\bmin{.48}
\centering\epsfig{file=./fig15b.ps,height=75mm,width=\linewidth,angle=270}
\emin
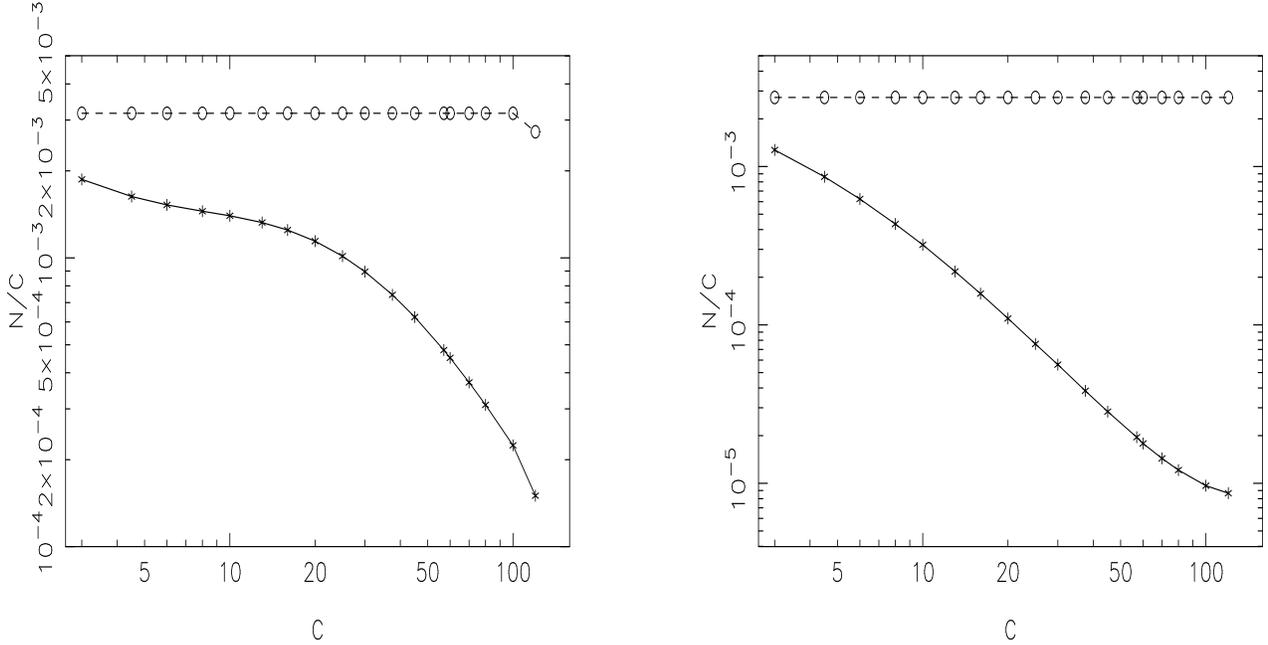
\caption{$N/C$ (solid line) and
$N^{\rm dp}/C$ (dashed line) as functions of $C$. All cases have 
$\omega=0.875$ and $R_{\rm ext} = 60$. {\it Left Panel:} Results with the
first outer boundary condition. The case with $C=120$ has $Nr=120$
while the rest of the cases have $Nr=100$. {\it Right Panel:} Results with the
second outer boundary condition. Many more models are employed here than
those tabulated in Table~\ref{table1}. The torque is seen to vary
smoothly with $C.$}  
\label{torquesc3t120of}
\efig

\section{Discussion}
\label{Discussion}

We first summarise our results. For a Keplerian rotation profile
 with the first boundary condition
the maximum inflation observed in our calculations
with $C=3$ is  such that $R_{\rm
f}/R_{\rm i} =3.2.$ This is a confirmation of the results of 
\citeN{bardou}. The expansion of the field lines is
restricted by the outer boundary in this case.
 However, for $|B_\varphi/B_z|
\sim 3$ at the disc surface the field lines with $r \le 3 R_{\rm c}$
do not vary much once $R_{\rm ext} > 10.$ Thus 
although significantly inflated they will
remain closed in the force--free equilibrium. Inflation increases
  with $C$ so that when $|B_\varphi/B_z|=10,$  a
maximum $R_{\rm f}/R_{\rm i} \sim 5$ is found for $R_{\rm i} =8.$

In calculations with $C=30$--$120$ and $R_{\rm ext}=11$ a
new effect was found. Inner field lines become deflated rather than inflated
with respect to the initial dipolar configuration. We expect the level
of deflation, and consequent increase
in poloidal magnetic pressure,
 to depend on the  rate of increase of toroidal 
magnetic pressure, which it has to balance around
$R=R_{\rm c}$. We  demonstrated, by varying
$R_{\rm ext}$, that the deflation
is not related to the outer boundary.
 We found that the level
of field line  deflation for $r < 10$ was restricted when
a smoother non Keplerian rotation profile was used.                     
 However, some level of deflation could be typical of the
field structure in the region  near the corotation radius. That
applies especially to discs with $\eta/\nu \sim 1$ and with $R_{\rm d}
\sim R_{\rm c}$ when a Dirichlet outer boundary condition is used.

In all the calculations presented here 
the  equilibrium poloidal field  is more nearly  radial
above the disc surface than the  stellar dipole field.
The lengthening of the field lines is accompanied by
a reduction of $|B_\varphi|/B_{\rm p}$ for increasing $\mu.$
We expect that  the force--free
  condition   constrains
 $|B_\varphi|/B_{\rm p}$  to remain  of order unity even for
$|B_\varphi/B_z| \gg 1$ at the disc surface. 

We also performed calculations with a Neumman outer boundary condition
that required the disc field to be radial at the outer boundary
as it might occur if a wind existed.
In these cases, other parameters
being equal,  the equilibrium field structure was more open.
The ratio $|B_\varphi| /B_{\rm p}$ was found to be less than unity for
$\mu < 0.1$ and $ \approx 0$ in most of the corona as
expected  on open field lines. In the limit of a large
twist the 
poloidal magnetic field 
tends to be expelled from the disc and consequently the toroidal
magnetic field is annihilated, as long as the disc field is permitted
to penetrate the outer boundary.                    

We  also found that for large
values of $|B_\varphi/B_z|$ on the disc surface the total  magnetic torque
 is {\it much
smaller} than that estimated  assuming $B_z \sim B_{\rm dp}$.
For
$C=120$  the torque is only
 $\sim 0.003$ of the dipole value
when the second boundary condition is used
 and $\sim 0.013-0.05$ times  the dipole value when
 the first boundary condition  is used (depending on the form of
$\Omega$). Since reasonable values of $\eta$ for circumstellar discs  tend
to produce very large
values of $|B_\varphi/B_z|$ on the disc surface, we believe that
the modification of the total torque  discussed here will  be
significant if a force free equilibrium can be attained. We therefore
expect significant modification of the magnetic breaking times
of T Tauri stars to result from the large twisting of the
magnetospheric field lines.             

\subsection{The spin--down of neutron stars}

Recent observations of accreting neutron stars
(\citeNP{nelson97}; \citeNP{chakra97}) suggest that
the  total torque  between star and disc
can oscillate producing alternating
spin--up and spin--down phases with little evidence of any correlation 
with the mass accretion rate.
\citeN{liw98} have suggested that this may be due
to variations in the structure of the magnetosphere.

The work presented here indicates that a large range in
the value
 of the spin--down torque may be obtained depending
on the outer  boundary condition and effective disc resistivity.
In cases favouring large field line inflation  and open field lines
the spin--down torque resulting from the disc-magnetosphere
interaction becomes very small. Thus if the magnetosphere
oscillates between such a state and one with smaller inflation,
oscillations in the spin--down torque and hence the total
torque may  be produced. The change from open to closed field lines 
may also affect the magnetic torque that arises from a wind, if
such a wind is produced. Changes in the magnetospheric
configuration might occur through reconnection through the
disc midplane, variations in outer conditions being more
or less favourable for open field configurations, or variations
in the internal disc resistivity.

\section*{Acknowledgments}
This work was supported by the European Union grant
ERBFMRX-CT98-0195. V.A. acknowledges support by the
State Scholarships Foundation (IKY) of the Hellenic Republic through
a postgraduate studentship. The authors are grateful to Anne Bardou
for useful discussions. 

\bibliography{mnrasmne,papers}
\bibliographystyle{mnras}

\end{document}